\documentclass[aps,prl,reprint,groupedaddress,twocolumn,superscriptaddress]{revtex4-1}
\usepackage{graphicx}     
\usepackage{subfigure}
\usepackage{dcolumn}     
\usepackage{bm}              
\usepackage{hyperref}     
\usepackage{amsmath}   
\usepackage{amssymb}
\usepackage{tabularx}
\usepackage{wasysym}
\usepackage{color}
\usepackage[table]{xcolor}
\usepackage{array}
\usepackage{multirow}
\usepackage{makecell}
\usepackage[version=3]{mhchem} 
\usepackage{float}
\usepackage{lipsum}
\usepackage{ulem}

\usepackage{verbatim}    
 
\makeatletter
\newcommand{\colorcaption}[2][]{%
  \begingroup%
  \renewcommand{\@caption@fignum@sep}{ (color online). }%
  \caption[#1]{#2}%
  \endgroup%
  }
\makeatother

\makeatletter
\newcommand*{\rom}[1]{\expandafter\@slowromancap\romannumeral #1@}
\makeatother

\begin{document}
\title{Critical magnetic fields and electron-pairing in magic-angle twisted bilayer graphene}
\author{Wei Qin}
\author{Bo Zou}
\author{Allan H. MacDonald}
\affiliation{Department of Physics, The University of Texas at Austin, Austin, Texas 78712,USA}
\email{macdpc@physics.utexas.edu}
\date{\today}

\begin{abstract}
The velocities of the quasiparticles that form Cooper pairs in a superconductor are revealed by the upper critical magnetic field. Here we use this property to assess superconductivity in magic angle twisted bilayer graphene (MATBG), which has been observed over a range of moir\'e band filling, twist angle, and screening environment conditions. We find that for pairing mechanisms that are unrelated to correlations within the MATBG flat bands, minima in an average Fermi velocity $v_F^* \equiv k_B T_c \ell_c /\hbar $, where $\ell_c$ is the magnetic length at the critical perpendicular magnetic field, are always coincident with transition temperature maxima. 
Both extrema occur near flat-band van Hove singularities. Since no such association is present in 
MATBG experimental data, we conclude that electronic correlations that yield a band-filling-dependent pairing glue must play a crucial role in MATBG superconductivity.
\end{abstract}

\maketitle

\textit{Introduction---} The observation of superconducting domes near correlated insulating states in 
magic-angle twisted bilayer graphene (MATBG) \cite{Cao:2018aa,Cao:2018ab} has stimulated interest in achieving a full microscopic 
understanding of this relatively simple electronic system \cite{Yankowitz:2019aa,Xie:2019aa,Kerelsky:2019aa,Choi:2019aa,Jiang:2019aa,Sharpe:2019aa,Lu:2019aa,Hu:2019aa,Serlin:2020aa,Julku:2020aa,Xie:2020aa,Wong:2020aa,Zondiner:2020aa,Balents:2020aa,Khalaf:2020aa,Stepanov:2020ab,Park:2020aa}. 
The electronic properties of MATBG devices are extremely sensitive to multiple tuning knobs, 
especially electrostatic doping and twist angle \cite{Cao:2018aa,Cao:2018ab,Lu:2019aa}, but also interlayer separation \cite{Carr:2018aa,Yankowitz:2019aa}, vertical displacement field \cite{Yankowitz:2019aa}, and three-dimensional screening environment \cite{Stepanov:2020aa,Saito:2020aa,Liu:2020aa}.
The tunability of MATBG makes it a particularly appealing experimental 
platform for the exploration of strong-correlation superconductivity.

As often happens in superconductors, observations that clearly distinguish between purely electronic 
pairing mechanisms, possibly related to the correlated insulating states \cite{Guo:2018aa,Dodaro:2018aa,Liu:2018aa,Isobe:2018aa,Fidrysiak:2018aa,Kennes:2018aa,Sherkunov:2018aa,Gonzalez:2019aa,Roy:2019aa,Huang:2019aa,Ray:2019aa,You:2019aa},
and conventional phonon-mediated electron pairing \cite{Wu:2018aa,Choi:2018aa,Lian:2019aa} are sparse.
In this Letter, we demonstrate by explicit calculations for representative flat-band and 
optical and acoustic phonon-mediated pairing models
that if superconductivity were mediated by a pairing glue derived from non-electronic degrees-of-freedom,
the maximum critical temperature would always occur when the Fermi level is close to the magic-angle flat-band 
van Hove singularity (VHS).  The band filling factor at which the VHS appears
in a particular system is extremely sensitive to electronic structure model details 
and also to sample-dependence, particularly in connection with broken flavor 
symmetries and associated Fermi surface reconstructions \cite{Xie:2020ab,Liu:2019aa,Liu:2019ab,Xie:2020ac}.  It is therefore
not easily predicted theoretically and is in all likelihood device dependent.
As we show here however, it can be determined experimentally 
by combining measured critical temperatures and critical perpendicular 
magnetic fields to determine the typical velocity of the quasiparticles that form Cooper pairs.  
Using this strategy we argue that there is no experimental correlation between a high transition temperature 
and van-Hove proximity, which argues against conventional phonon-mediated 
pairing.

\begin{table}
\caption{Experimental results for the critical temperature $T_c$, critical perpendicular magnetic field $ H_{c2}$, Pauli critical field $H_P$, and the extracted average Fermi velocity $v_F^*$ for MATBG in a variety of experiments with different twist angles $\theta$, band fillings $\nu$, and screening environments. }
\label{tab:table1} 
\begin{ruledtabular}
\begin{tabular}{c c c c c c c}
     $\theta$ ($^\circ $)  &  $\nu$ & $T_c$ (K) & $H_{c2}$ (mT) & $H_P$ (T) & $v_F^*$ ($10^4$m/s) & Refs \\
     \hline
     1.05 &  -2.02 & 1.7  & - &  3.15 & - & \cite{Cao:2018aa}\\
     1.16  &  -2.15 & 0.5 &125 & 0.93   & 0.34 &\\
     \hline
     1.27  & -2.33 & 3.1  &210 &  5.74  &  1.54 & \cite{Yankowitz:2019aa}\\
     1.27  & -2.62 & 2 & 72 &  3.7   & 1.78 &   \\
     
     \hline
      & -2.31 &  3.1  & 180 & 5.74  &1.67&  \cite{Lu:2019aa} \\
  1.1    & -1  &  0.14  &100  &0.26  &  0.10& \\
              & 0.67 &  0.16 &  300 & 0.3 &  0.07 &\\
             & 1.48 &  0.65 &400 & 1.2&  0.23& \\
      \hline
      1.15   & -1.6  &  0.92 &220 & 1.7 &  0.46&\cite{Stepanov:2020aa} \\  
      	        & 1.8  &  0.42 &26 & 0.78 & 0.56 & \\
      \hline
     1.04    & -2.43  &  1.3 &$>$50 &2.41  & $<$1.38&\cite{Saito:2020aa} \\
     1.09   & -2.79  &  2.5 &45 & 4.63 &2.78  &\\
     1.18   & -2.5  &  0.7 &$>$60 & 1.3 & $<$0.68& \\
     1.12   & -2.47  &  4 &$>$50 & 7.4  & $<$4.24& \\
\end{tabular}
\end{ruledtabular}
\end{table}

Table~\ref{tab:table1} summarizes experimental data for MATBG superconductors. Each observation determines an average Fermi velocity defined by $v_F^* \equiv k_B T_c / \hbar q_c$, where $q_c=1/\ell_c$ is the maximum pairing momentum \cite{SI}, $2 \pi \ell_c^2 = \Phi_0/H_{c2}$, 
and the superconducting flux quantum $\Phi_0=2e/hc$. The average Fermi velocity defined 
above is motivated by the observation that superconductivity is suppressed at finite pairing momentum because 
the electrons that form Cooper pairs differ in energy by $\sim (d\epsilon_{k}/dk)/\ell \sim \hbar v_F / \ell$, 
and that it is lost in mean-field theory when this difference is comparable to $k_B T_c$.  
Microscopically, superconductivity is lost in a magnetic field because Landau level Cooper pair
states are formed from individual electron states
that differ in energy by $\sim \hbar v_F / \ell_c$ \cite{MacDonald:1993aa}.

\begin{figure}
 \centering
\includegraphics[width=\columnwidth]{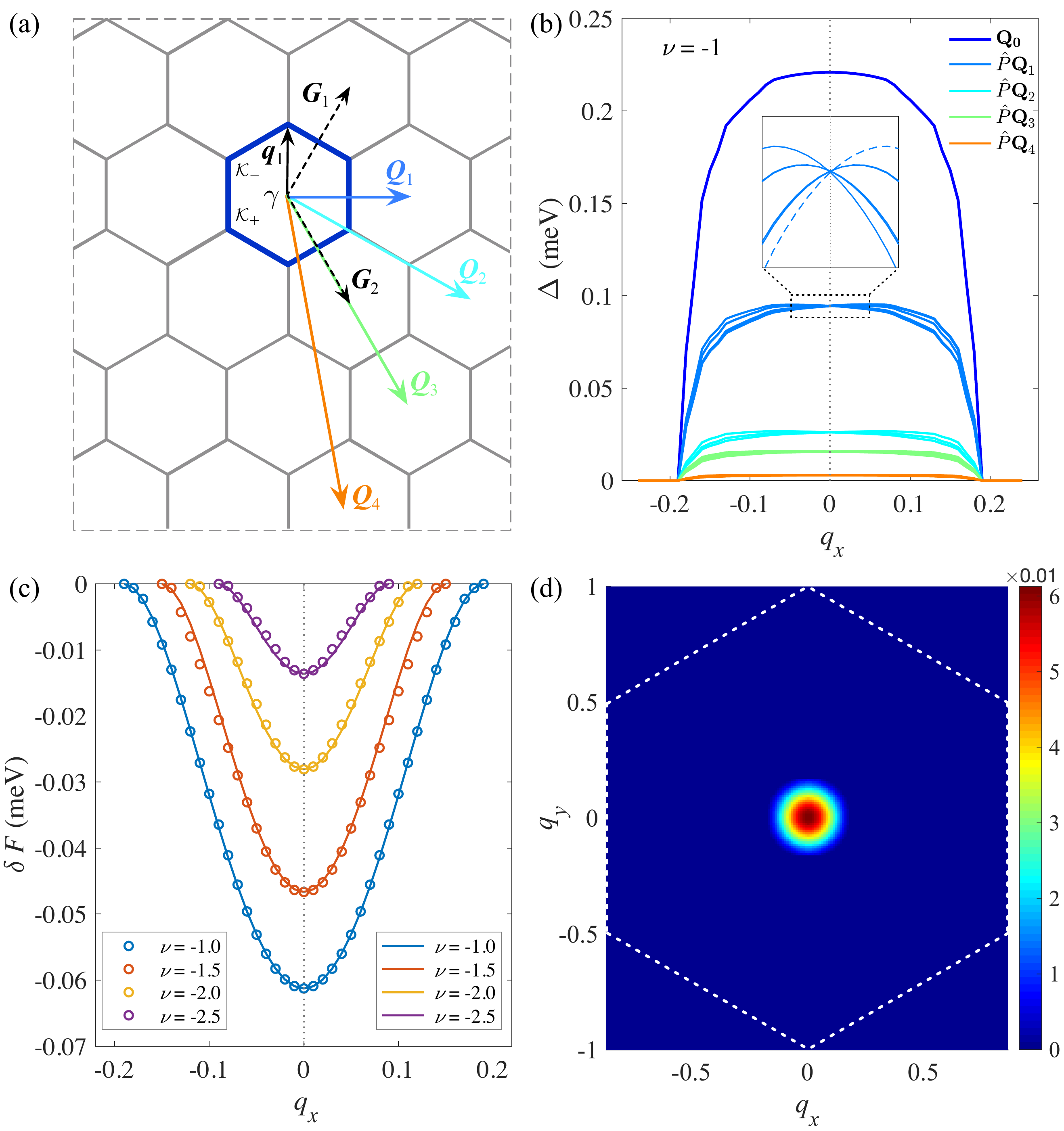}
 \caption{Pairing wavevector dependence of superconductivity in MATBG at $\theta = 1.07^{\circ}$. 
 (a) Schematic diagram of a moir\'e Brillouin zone (MBZ). (b) Fourier coefficients of the real-space pair potential expanded up to four shells of reciprocal lattice vectors, which are generated by acting symmetry operations $\hat{P}$ from point group $\mathcal{C}_6$ on $\bm{Q}_{i}$ depicted in (a), where $\bm{Q}_0 = 0$. The insert shows $\Delta_{\hat{P}\bm{Q}_1}$ for flat-band filling $\nu=-1$ at small pairing wavevector $q_x$, with the dashed curve corresponding to $\Delta_{\bm{Q}_1}$.
(c) Superconducting condensation energy normalized per moir\'e supercell {\it vs.} pairing wavevector $q_x$ for different band fillings, where circles are numerical results and solid curves 
are fits to Eq.~(\ref{eq:GLfeng}). (d) Color scale plot of $-\delta F(\bm{q})$ at $\nu = -1$, where the MBZ is indicated by the dashed hexagon. 
These results are obtained at zero temperature with the interlayer tunneling ratio set to $\eta = t_{\text{AA}}/t_{\text{AB}} = 0.85$, and Coulomb repulsion $u = 40$ meV$\cdot$nm$^2$
chosen to yield $T_c \sim 1.7$ K.} 
 \label{fig:figure1}
\end{figure}

\begin{figure}
  \centering
 \includegraphics[width= \columnwidth ]{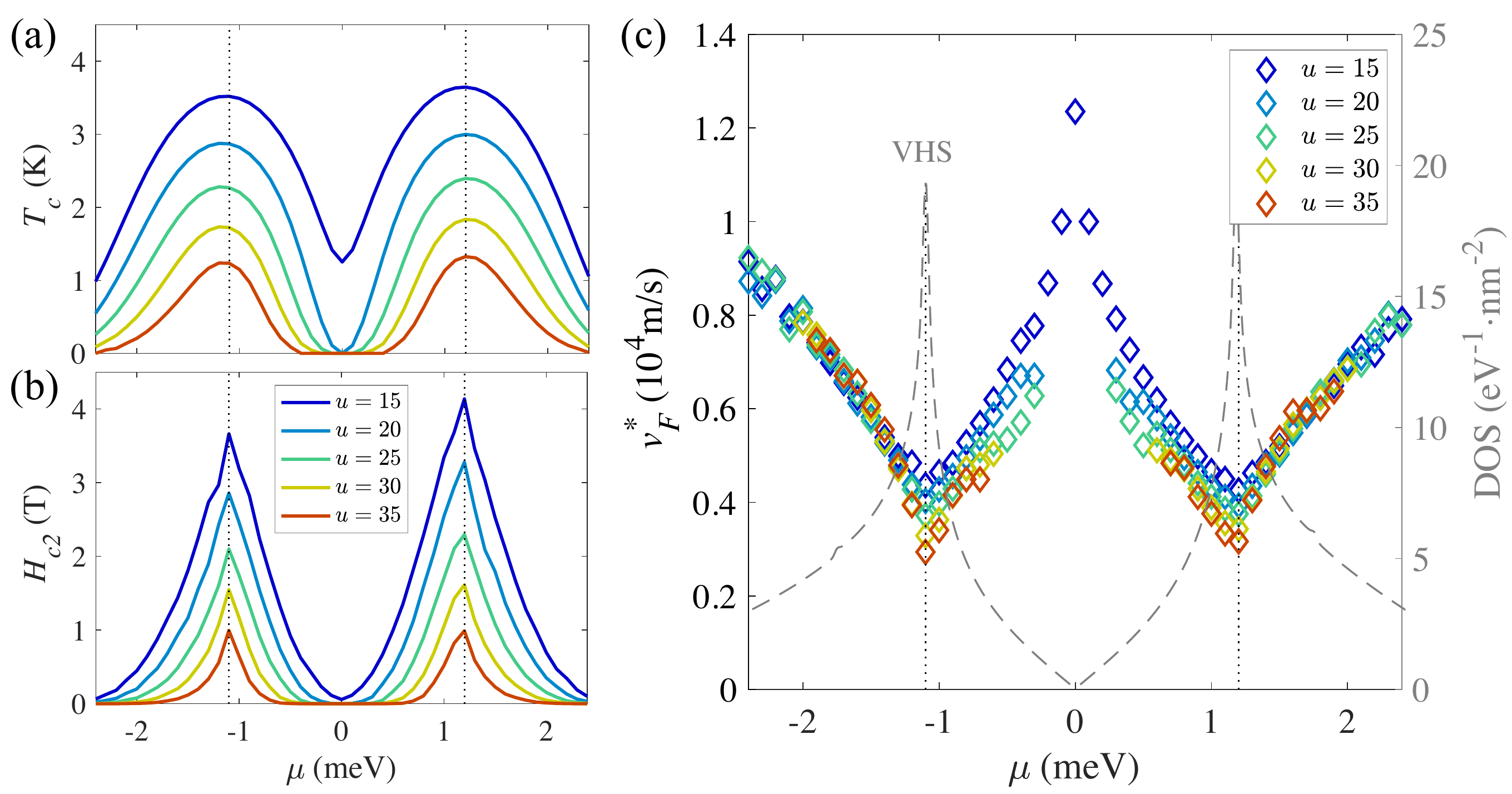}
  \caption{(a)-(c) Zero-pairing-momentum critical temperature $T_c$, perpendicular critical magnetic field $H_{c2}$, and average Fermi velocity $v_F^*$ as functions of chemical potential $\mu$ for several strengths of Coulomb repulsion 
  $u$ in units of meV$\cdot$nm$^2$.
  These results are calculated for a MATBG band-structure model with $\theta = 1.07^{\circ}$ and $\eta =0.7$. 
  The dotted lines indicate the values of $\mu$ for flat-band VHSs in the DOS, which is plotted as a dashed curve in (c). $v_F^*$ in (c) has a minimum where $T_c$ in (a) is maximized.  
    } 
  \label{fig:figure2}
\end{figure}

We will show by explicit calculation that this 
experimentally available average Fermi velocity very effectively identifies VHS 
in two-dimensional band structures.  We note that the experimental values of $v_F^*$
in Table~\ref{tab:table1} are typically 
a hundred or more times smaller than the Dirac velocities of isolated graphene sheets, 
demonstrating the crucial role of the dramatically flattened moir\'e bands \cite{Bistritzer:2011aa}. The main point we wish to make here however is that experimental critical temperature maxima, which always occur in a narrow range of filling factor near $\nu=-2.3$ \cite{Cao:2018aa,Yankowitz:2019aa,Lu:2019aa,Saito:2020aa,Stepanov:2020ab},
do not correlate with average Fermi velocity minima as they would in any theory in which 
pairing is mediated by phonons, or other bosons that are insensitive to $\nu$.
Below we first describe microscopic calculations of $T_c$ and $H_{c2}$ based on a 
phonon-mediated pairing model that demonstrate this point, and then discuss the implications 
of this observation.  Our main results are summarized in Fig.~\ref{fig:figure1} and Fig.~\ref{fig:figure2}.

\textit{Critical magnetic field from finite-momentum pairing---}
The conclusions in this paper are based on calculations of the pairing wavevector 
dependence of the condensation 
energy for MATBG superconductors. We assume that superconducting 
condensation energy can be calculated using a mean-field approximation, and that the relevant Cooper pairs involve two electrons from opposite valleys.  Given these assumptions, the theoretical system properties depend only
on the MATBG band structure model and the interaction Hamiltonian.  The former is not accurately 
known at present, mainly because bands are renormalized by 
interactions \cite{Xie:2020ab,Liu:2019aa,Liu:2019ab}, and because these renormalizations are sensitive to the 
three-dimensional screening environment \cite{Goodwin:2019aa,Goodwin:2020aa}.  We will account for the possible role of band-structure variations 
by varying $\eta$, the ratio between the intrasublattice and intersublattice tunneling terms in the
Bistritzer-MacDonald MATBG model \cite{Bistritzer:2011aa}.

In the phenomenological Ginzburg Landau theory of
an isotropic superconductor \cite{SI}, the superconducting condensation energy 
at finite pairing-momentum
\begin{equation}
\delta F(\bm{q}) = \delta F_0 [1-\left(\bm{q}/q_c\right)^2 ]^2,
\label{eq:GLfeng}
\end{equation}
where $\delta F_0$ is the condensation energy at zero pairing-momentum
and $q_c$ is the critical pairing wavevector to which superconductivity survives.
Our microscopic calculations are in close agreement with this expression and can be fit to 
determine $\delta F_0$ and $q_c$.
As explained in the Supplemental Material (SM) \cite{SI}, the critical perpendicular magnetic field  
is related to $q_c$ by $H_{c2} = \Phi_s q_c^2/2\pi$. 
In addition, the supercurrent density as a function of pairing wavevector can also be calculated from  $\delta F(\bm{q})$ using $\bm{j} = (2e/\hbar) [\partial F(\bm{q})/\partial \bm{q}]$ \cite{SI}.

\textit{Mean-field theory for finite-momentum pairing---}
The finite pairing-momentum mean-field calculations that we have performed 
for MATBG superconductors employ models in which electrons are paired by  
in-plane optical phonons \cite{Wu:2018aa,Choi:2018aa,SI} competing with repulsive 
Coulomb interactions.  Because the flat band width is small compared to the phonon energy, both 
interactions are essentially instantaneous. Optical phonons are a convenient choice for a putative pairing mechanism because their interactions with graphene $\pi$-bands are well understood \cite{Wu:2018aa,Basko:2008aa}.  The case of acoustic-phonon mediated interactions is 
discussed later.

The long-range Coulomb interaction is thought to be responsible for 
flavor-symmetry-breaking, also discussed later, which plays an important 
role in MATBG physics \cite{Xie:2020ab,Liu:2019aa,Liu:2019ab}.  
Here we assume that there is negligible Coulomb scattering between 
valleys and adjust the strength of repulsive intra-valley Coulomb scattering,
which plays a depairing role, to match experimental critical temperature.
(This adjustment however sets the Coulomb scale to value that is smaller than the expectation from a crude estimation. Coulomb repulsion is assumed to have the 
same strength for electrons from the same valley and opposite valleys.)
Specifically, the model electron-electron interaction Hamiltonian $
H_{ee} = 2u\sum_{ll'} \sum_{\tau s \sigma \sigma' } \int d\bm{r} \rho_{l \tau s \sigma}(\bm{r}) \rho_{l' \bar{\tau} \bar{s} \sigma' }(\bm{r}),
$ where the density operator $\rho_{l \tau s \sigma }(\bm{r})  = \psi^{\dagger}_{l \tau s \sigma}(\bm{r}) \psi_{l 
\tau s \sigma}(\bm{r})$, $\psi_{l 
\tau s \sigma}(\bm{r}) $ is the real-space annihilation field operator, 
and $l,\tau, s, \sigma = \pm1$ specify
layer, valley, spin and sublattice.

The self-consistent calculations for the 
mean-field superconducting state at finite pairing momentum are standard and briefly
summarized in the following paragraph. The finite-momentum pairing MATBG Bogoliubov-de Gennes (BdG) Hamiltonian \cite{SI}
\begin{equation}
\mathcal{H}_{\text{BdG}}(\bm{q},\bm{k}) = 
\begin{bmatrix}
\mathcal{H}(\bm{k}) -\mu_{\bm{q}} & \Delta_{\bm{q}}(\bm{k}) \\
\Delta^{\dagger}_{\bm{q}}(\bm{k}) & -\mathcal{H}^{\text{T}}(\bm{q}-\bm{k}) +\mu_{\bm{q}},
\end{bmatrix},
\label{eq:BdG}
\end{equation}
where each block acts on four-component sublattice spinors and the pair potential for pairing wavevector $\bm{q}$
\begin{equation}
[\Delta_{\bm{q}} (\bm{k})]_{\alpha \alpha_1} = \sum_{\alpha' \alpha_1'}  \sum_{\bm{k}'} [V(\bm{k},\bm{k}')]_{\alpha  \alpha_1\alpha_1' \alpha'} [\mathcal{F}_{\bm{q}}(\bm{k}')]_{\alpha' \alpha_1'}. 
\label{eq:gap}
\end{equation}
Here $V(\bm{k},\bm{k}')$ is the total interaction matrix element including phonon-mediated attraction and Coulomb repulsion contributions, $\mathcal{F}_{\bm{q}}(\bm{k}')$ is Gorkov's anomalous Green's function \cite{Mahan:2013aa}, and the summation of $\bm{k}'$ is over the MBZ. 
The chemical potential $\mu_{\bm{q}}$ is determined self-consistently by particle number conservation. The free energy of the finite pairing-momentum superconducting state
\begin{equation}
F_s(\bm{q}) =   C_{\bm{q}} + A n_0 \mu_{\bm{q}} + \frac{1}{2\beta} \text{Tr}  \sum_{\bm{k}} \ln f[- E_{\bm{q}}(\bm{k})], 
\label{eq:feng}
\end{equation}
where $n_0$ is the carrier density measured from charge neutrality, 
$A$ is the sample area, $f(\epsilon)$ is the Fermi-Dirac distribution function, $E_{\bm{q}}(\bm{k})$ are the eigenvalues of the BdG Hamiltonian, and $C_{\bm{q}} =- \frac{1}{2}\text{Tr}  \mathcal{F}^{\dagger}_{\bm{q}} V \mathcal{F}_{\bm{q}}$ is a double-counting correction \cite{SI}. 
The superconducting condensation energy is defined as 
$\delta F(\bm{q}) = F_s(\bm{q})-F_n$, where $F_n$ is the normal-state free energy calculated by 
Eq.~(\ref{eq:feng}) for vanished pair potential.  We use numerical results for the 
condensation energy as a function of pairing wavevector, band filling $\nu$,
and model parameters to connect with experimental observables.

\textit{Robust correlations between velocity minima and critical temperature maxima--} 
Because our model interaction is local the real-space pair potential is conveniently paramaterized by performing a Fourier expansion.  The coefficients of reciprocal lattice
vectors in this expansion are plotted as a function of pairing wavevector in the MBZ
in Fig.~\ref{fig:figure1}(b). In the insert of Fig.~\ref{fig:figure1}(b), the asymmetric behavior of $\Delta_{\mathcal{P} \bm{Q}_1}$ is due to rotational symmetry breaking at finite-momentum pairing. 
Figures~\ref{fig:figure1}(c) and (d) depict 
the pairing wavevector dependence of $\delta F(\bm{q})$, which possesses a minimum at $\bm{q} = 0$, indicating that 
the zero-momentum pairing state is the ground superconducting state. 
Moreover, $\delta F(\bm{q})$ is nearly isotropic and vanishes at a critical pairing wavevector $q_c$. The small value of $q_c$ compared to the 
MBZ implies that the superconducting coherence length is many times of the 
moir\'e period.

Figures \ref{fig:figure2} (a)-(b) show mean-field results for the 
zero-pairing-momentum critical temperature $T_c$, and the critical perpendicular magnetic field $H_{c2}$ 
obtained from finite-pairing-momentum calculations.
As expected, $T_c$ decreases monotonously upon strengthening the repulsive Coulomb interaction between electrons. Dome-like features are revealed in $T_c$ as a function of $\mu$ which are 
reminiscent with the superconducting domes, but are associated with maxima centered on the flat-band VHSs. 
$H_{c2}$ possesses sharp peaks at VHSs consistent with the phenomenological argument that $H_{c2}$ is proportional to the effective mass of paired electrons \cite{SI}. 
Away from the VHSs, $H_{c2}$ decreases quickly and approaches zero when $T_c$ becomes zero. 
Figure \ref{fig:figure2}(c) shows the average Fermi velocity extracted from the mean-field results of 
$T_c$ and $H_{c2}$ via $v_F^* = k_BT_c/\hbar q_c$.  The average velocities are a hundred or more times smaller than the Dirac velocity of isolated graphene sheets
$v_D \sim10^6$ m/s, in agreement with experiment.  In contrast to experiment, $v_F^*$ possesses minima at the VHSs and therefore at the $T_c$ maxima. 

\begin{figure}
  \centering
 \includegraphics[width=  \columnwidth ]{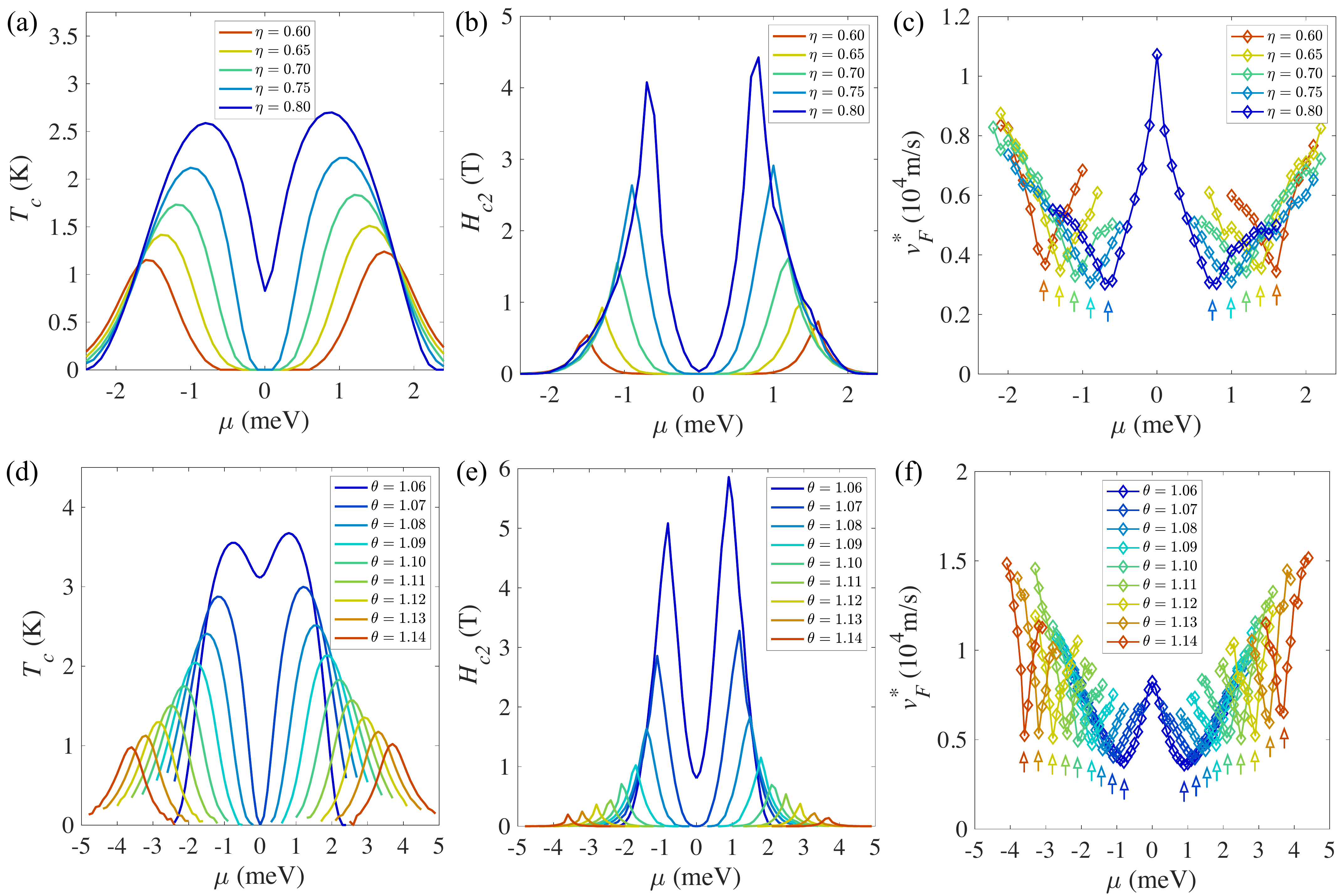}
  \caption{ (a)-(c) $T_c$, $H_{c2}$, and $v_F^*$ as functions of $\mu$ for several values of 
  $\eta$ for MATBG with $\theta = 1.07^{\circ}$ and $u=30$ meV$\cdot$nm$^2$. 
  (d)-(f) $T_c$, $H_{c2}$, and $v_F^*$ as functions of $\mu$ for several different twist angles for MATBG with $\eta = 0.7 $ and $u=20$ meV$\cdot$nm$^2$. Vertical arrows in (c) and (f) highlight the positions of flat-band VHSs.} 
  \label{fig:figure3}
\end{figure}

In Fig.~\ref{fig:figure2}(c), we have seen that the average Fermi velocity is
almost perfectly a band structure property that is independent of the pairing strength. 
In Fig.~\ref{fig:figure3}, we study the correlation between $T_c$ and $v_F^*$ by varying the 
band-structure parameters $\eta$ and $\theta$ for fixed pairing strength.
Over the illustrated parameter range the flat-band width is larger at smaller $\eta$ and larger $\theta$ \cite{SI}. 
Accordingly decreasing $\eta$ or increasing $\theta$ tends to reduce $T_c$ as shown in Fig.~\ref{fig:figure3} (a) and (d), where dome-like features in $T_c$ are peaked around the shifting VHSs. 
As illustrated in Fig.~\ref{fig:figure3} (b) and (e), sharp peaks of $H_{c2}$ are also found at the VHSs, and the magnitude of $H_{c2}$ is dramatically suppressed for models with larger flat-band width. The extracted $v_F^*$ are plotted in 
Fig.~\ref{fig:figure3} (c) and (f), and are as expected larger for larger flat-band widths.
For every band-structure model, $v_F^*$ possesses a V-shaped minimum near the VHS, which is always coincident with the maximum of $T_c$.

\begin{figure}
  \centering
 \includegraphics[width= \columnwidth ]{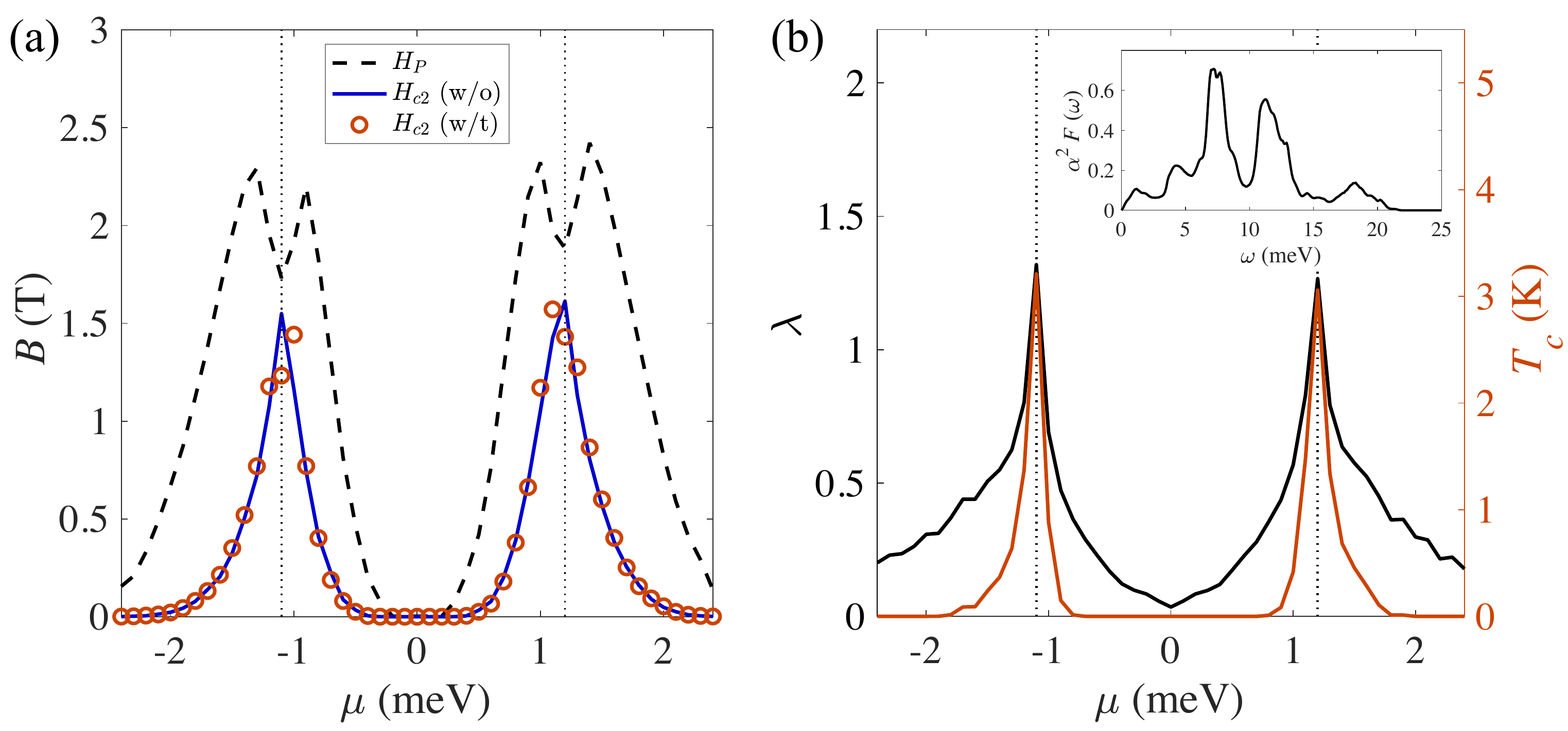}
  \caption{(a) Critical magnetic field calculated at zero pairing wavevector by including 
  Zeeman coupling ($H_P$), and calculated from the critical pairing wavevector $H_{c2}$ without (w/o) and with (w/t) Zeeman coupling. In the presence of Zeeman coupling, $H_{c2} = (\Phi_s/2\pi) q_{x,c} q_{y,c} $,  where $q_{x,c}$ and $q_{y,c}$ are the critical pairing wavevectors along $x$ and $y$ directions. These results are for $\theta = 1.07^{\circ}$, $\eta = 0.7$, and $u = 30$ meV$\cdot$nm$^2$. 
  (b) Electron-acoustic phonon coupling strength $\lambda$ and the estimated $T_c$ for MATBG with $\theta = 1.07^{\circ}$ and $\eta = 0.7$. The insert gives the Fermi-surface averaged electron-acoustic phonon spectral function $\alpha^2 F(\omega)$ at $\mu = -1.1$ meV, i.e. the VHS of the valence flat band.
  The dotted vertical lines in (a) and (b) show the positions of the VHSs. } 
  \label{fig:figure4}
\end{figure}

\textit{Discussion---}
So far we have neglected both the possibility of flavor symmetry breaking and 
the role of Zeeman coupling to the electronic spin.  Indeed, this assumption can be 
questioned since there is strong experimental evidence that the strongest superconducting dome occurs in a region of filling factor where only two flavors are partially occupied and the 
moir\'e flat-bands have consequently reconstructed \cite{Cao:2018aa,Cao:2018ab,Yankowitz:2019aa,Xie:2019aa,Kerelsky:2019aa,Choi:2019aa,Jiang:2019aa,Sharpe:2019aa,Lu:2019aa,Serlin:2020aa,Wong:2020aa,Zondiner:2020aa, Xie:2020ab,Liu:2019aa,Liu:2019ab,Xie:2020ac}. 
If we were to assume that the superconducting state 
near $\nu=-2.3$ is spin-polarized, with partially occupied valence bands for two different valleys with the same spin,
the neglect of Zeeman coupling would be appropriate. The only difference between the spin-polarized and 
spin-unpolarized calculations then, would be a change in how the intervalley electron-phonon interactions enter the gap equation \cite{SI}, leading simply to a change in the effective 
pairing strength which would not alter our conclusions.
On the other hand, if the superconducting state is spin-singlet, the pair-breaking effect of Zeeman coupling 
would need to be considered. In Table \ref{tab:table1}, we have listed Pauli critical fields extracted from experimental data by using $H_P \approx 1.85 T_c$ (T) \cite{Cao:2018aa}, which is much larger than $H_{c2}$ for large $T_c$, indicating the orbital effect dominates pair-breaking. 
Figure~\ref{fig:figure4}(a) compares critical magnetic fields calculated by including 
only the Zeeman effect, only the orbital effect, and both effects.  
Close to the VHS, $H_P$ is comparable to $H_{c2}$ because the 
orbit effect is suppressed due to small Fermi velocities.
Away from the VHSs, however, $H_P$ is much larger than $H_{c2}$, consistent with experimental observations. 
The two peaks in $H_P$ near the VHS reflect Zeeman splitting. 
The critical fields calculated including both effects show that $H_{c2}$ 
is nearly unchanged by Zeeman coupling except when the Fermi level is very close to the VHS. 

We have so far not explicitly included acoustic-phonon mediated interactions, which may compete 
more successfully with Coulomb interactions because they are retarded.
The longitudinal acoustic phonon velocity in an isolated graphene sheet 
$v_{ph} = 2\times10^4$ m/s is comparable to the flat-band electron velocities.
Figure~\ref{fig:figure4}(b) illustrates the band-filling dependence of 
the dimensionless electron-acoustic-phonon coupling constant $\lambda = 2\int d\omega \alpha^2F(\omega)/\omega$
and the non-trivial frequency dependence of the Fermi-surface averaged coupling strength $\alpha^2F(\omega)$
for a representative band-structure model \cite{SI}.  
The critical temperatures shown in Fig.~\ref{fig:figure4}(b) are estimated by using the formula 
$T_c = \frac{\hbar \omega_{\text{ln}}}{1.2k_B} \exp{[\frac{1.04(1+\lambda)}{\lambda-\mu^*(1+0.62\lambda)}]}$, 
where the averaged phonon frequency $\omega_{\text{ln}} = \exp{[ 2 \int d \omega \ln{(\omega)}  \alpha^2F(\omega) /(\lambda \omega)]} $ \cite{Allen:1975aa}, and the reduced Coulomb coupling strength $\mu^* = 0.2$.
In contrast to the optical-phonon-mediated interaction, retardation does supply a formal justification
for reduced Coulomb coupling \cite{Morel:1962aa}, but since the phonon and electronic energy scales are similar, 
it still does not justify the large reduction needed to match experimental $T_c$ scales.
That aside, it is clear in Fig.~\ref{fig:figure4}(b) that the association of $T_c$ maxima with 
flat-band VHSs applies equally well to acoustic-phonon mediated superconductivity.

For a given pairing glue, weak-coupling BCS theory predicts that $T_c$ is positively correlated with the DOS at Fermi level, determined by the 
Fermi surface size and the typical quasiparticle velocity.
In experiment \cite{Cao:2018aa,Cao:2018ab,Park:2020aa}, these quantities are often extracted from 
the frequency and temperature dependence of magnetic oscillations, which respectively measure 
Fermi surface area and cyclotron effective masses.  These measurements are 
normally carried out in the high magnetic field regime, where superconductivity in MATBG is completely suppressed. 
In this study, we propose extracting average Fermi velocity instead from a combination of the 
$T_c$ and $H_{c2}$, which is more directly 
connected to the properties of quasiparticles that participate in pairing. We establish the reliability 
of this approach by calculating $T_c$ and $H_{c2}$ for a wide variety of representative MATBG models.
In particular, our theoretical calculations show that the definition of $v_F^*$ in this way are largely independent of 
pairing model, that they exhibit a V-shaped cusp at the flat-band VHS, and that that $v_F^*$ is always negatively 
correlated with $T_c$. Since experimental values of $v_F^*$,
summarized in Table I and in the SM \cite{SI}, do not show any indication of a such correlation between $v_F^*$ and $T_c$, we conclude that 
the pairing glue is related to electronic fluctuations that are sensitive to flat-band filling factor
and optimized near the peaks of the experimental superconducting domes.

\textit{Acknowledgment---} The authors acknowledge helpful discussions with Andrea Young and Ming Xie. This work was supported by DOE grant DE-FG02-02ER45958 and Welch Foundation grant TBF1473

\end{document}


\title{Supplemental Material: Critical magnetic fields and electron-pairing in magic-angle twisted bilayer graphene}
\author{Wei Qin}
\author{Bo Zou}
\author{Allan H. MacDonald}
\affiliation{Department of Physics, The University of Texas at Austin, Austin, Texas 78712,USA}
\email{macd@physics.utexas.edu}
\date{\today}

\maketitle

\section{continuum model of MATBG}
\begin{figure}
\centering
\includegraphics[width=\columnwidth]{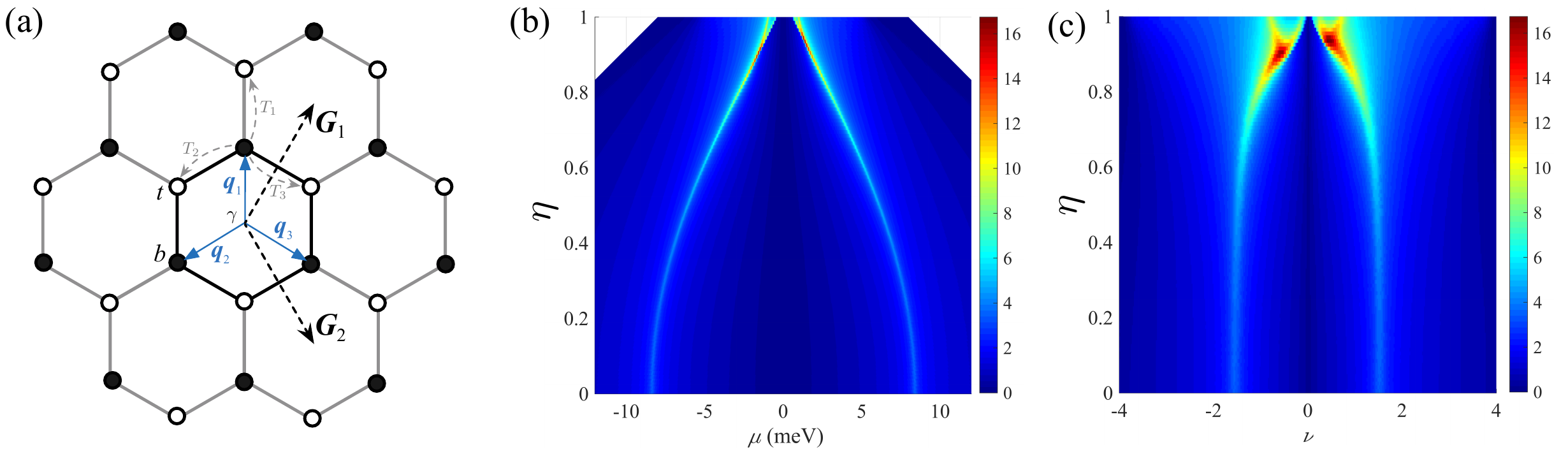}
\caption{(a) Momentum-space honeycomb lattice spanned by vectors $\bm{q}_{1,2,3}$, where the black solid hexagon denotes the first moir\'e Brillouin zone (MBZ). The filled and hollow circles represent Dirac cones from the bottom and top layers of graphene, which are coupled through nearest neighbor tunneling matrix $T_{1,2,3}$ as schematically depicted by the gray dashed arrows. $\bm{G}_{1,2}$ are the primitive moir\'e reciprocal lattice vectors. (b)-(c) Color scale plot of the density of states of MATBG with $\theta = 1.15^{\circ}$ {\it vs.} $\eta = t_{\text{AA}}/t_{\text{BB}}$ and (b) chemical potential $\mu$, and (c) band filling $\nu$. The unit of color bar is eV$^{-1}\cdot$nm$^2$. }
\label{fig:figs1}
\end{figure}

In the present study, the moir\'e bands of magic angle twisted bilayer graphene (MATBG) are calculated within a continuum-model Hamiltonian \cite{Bistritzer:2011aa,Wu:2018aa}, where Dirac electrons at the two Brillouin zone corners 
of each graphene layer coupled by periodic sublattice-dependent local tunneling. The K-valley projected Hamiltonian
\begin{equation}
H_{0,K} = \begin{bmatrix}
-i\nu_F \bm{\sigma}_{-\theta/2} \cdot \bm{\nabla}  & T(\bm{r}) \\
T^{\dagger}(\bm{r}) & -i\nu_F \bm{\sigma}_{\theta/2} \cdot \bm{\nabla} 
\end{bmatrix},
\label{eq:KvH0}
\end{equation}
where $\bm{\sigma}_{\theta} = e^{i\theta/2 \sigma_z}(\sigma_x,\sigma_y)e^{-i\theta/2\sigma_z}$, $\sigma_{x,y,z}$ are Pauli matrices acting on sublattice. The periodic interlayer local tunneling $T(\bm{r}) = \sum_{j=1}^{3} T_j e^{i\bm{q}_j\cdot \bm{r}} $, where
\begin{equation}
T_j = t_{\text{AA}} \sigma_0 + t_{\text{AB}} [\cos(j2\pi/3)\sigma_x+\sin(j2\pi/3)\sigma_y].
\end{equation}
where $\bm{q}_1 = k_{\theta}(0, 1)$, $\bm{q}_{2,3} = k_{\theta}(\pm\sqrt{3}/2, -1/2)$, as illustrated in Fig.~\ref{fig:figs1}. The edge length of the moir\'e Brillouin zone $k_{\theta} = (8\pi/3a_0) \sin(\theta/2)$ with graphene lattice constant $a_0 = 2.46$ \text{\AA}. In the main text, we take $t_{\text{AB}} = 117$ meV and $ t_{\text{AA}} = \eta t_{\text{AB}} $ with $\eta <1$ accounting for corrugation and strain effects \cite{Carr:2019aa}. In the moir\'e momentum representation, the moir\'e Hamiltonian
\begin{equation}
H_{0} = \sum_{\alpha \alpha'}\sum_{\bm{k} \in \text{MBZ}} \psi_{\alpha}^{\dagger}(\bm{k}) \left[ \mathcal{H}_{\alpha \alpha'}(\bm{k})-\mu \delta_{\alpha \alpha'} \right]\psi_{\alpha'}(\bm{k}),
\end{equation}
where $\psi_{\alpha}(\bm{k}) $ is an annihilation field operator and 
$\alpha = (l \tau s \sigma n)$ is a lumped label with $l,\tau, s, \sigma = \pm1$ specifying
layer, valley, spin and sublattice, and $n = (n_1,n_2)$ specifying the 
moir\'e reciprocal lattice vector $\bm{G}_{n} = n_1 \bm{G}_{1} + n_2 \bm{G}_{2}$
with $n_{1,2}$ being integers. As depicted in Fig.~\ref{fig:figs1}, $\bm{G}_{1,2}= k_{\theta}(\sqrt{3}/2, \pm 3/2) $ are primitive moir\'e reciprocal lattice vectors. In the momentum space, the moir\'e Hamiltonian can be viewed as Dirac cones arranged on a honeycomb lattice and coupled via nearest-neighbor tunneling matrices, as schematically depicted in Fig.~\ref{fig:figs1}. Therefore, the matrix element of K-valley projected moir\'e Hamiltonian
\begin{equation}
[\mathcal{H}_{\text{K}}(\bm{k})]_{l\sigma n,l'\sigma' n'} = \hbar \nu_{F} e^{il \theta/4 \sigma_z } \bm{\sigma} \cdot [\bm{k}+l(\bm{G}_{n}-\bm{q}_1)] e^{-il \theta/4\sigma_z} \delta_{ll'}\delta_{\sigma \sigma'} \delta_{nn'}+\sum_{j =1}^{3} T_{j}\delta_{l,-l'} \delta_{\sigma \sigma'} \delta_{\bm{q}_{j}+2\bm{q}_1,\bm{G}_{n+n'}},
\end{equation}
where $v_F \sim 10^6$ m/s is the Fermi velocity of Dirac electron in isolated graphene. The two valleys in the continuum model of MATBG are decoupled, indicating $\mathcal{H}(\bm{k})=  \text{diag}[\mathcal{H}_{\text{K}}(\bm{k}),\mathcal{H}_{-\text{K}}(\bm{k})]$, where $\mathcal{H}_{-\text{K}}(\bm{k})= \mathcal{T}\mathcal{H}_{\text{K}}(\bm{k})$ with $\mathcal{T}$ denoting the time-reversal operator \cite{Bistritzer:2011aa}. 

Figure~\ref{fig:figs1}(b) and (c) illustrate a typical calculation of the density of states at different chemical potentials or band fillings upon varying the interlayer tunneling ratio $\eta = t_{\text{AA}}/t_{\text{AB}}$ for MATBG with $\theta = 1.15^{\circ}$. When decrease $\eta$, the flat-band width becomes larger and the band fillings of conduction and valence flat-band van Hove singularities approach $\nu = \pm2$, respectively.

\section{Ginzburg-Landau theory of superconductivity }
The Ginzburg-Landau (GL) theory of superconductivity is based on the expansion of free energy of a system in powers of superconducting order parameter \cite{Bennemann:2008aa}. In the presence of magnetic field the free energy 
\begin{equation}
F_s = F_n + \int d\bm{r} \left[ \frac{\hbar^2}{2m^*}|(\bm{\nabla}-i\frac{2e}{\hbar c}\bm{A})\psi|^2 + \alpha(T)|\psi|^2 +\frac{\beta(T)}{2} |\psi|^4 +\frac{\bm{B}^2}{8\pi} \right],
\label{eq:GLfeng}
\end{equation}
here $\psi = \sqrt{n_s} e^{i\phi} $ is the complex order parameter, $\phi$ is the phase of order parameter, $n_s$, $m^*$, and $2e$ are the density, effective mass, and total charge of Cooper pair. On the right hand side of Eq.~\ref{eq:GLfeng}, the first term in the integral is responsible for the kinetic energy of Cooper pair, the second term is the pairing potential energy, the third term resembles the two-body interaction energy, and the last term is magnetic energy. By varying the GL free energy with respect to magnetic vector potential $\bm{A}$ and order parameter $\psi^*$, the supercurrent density
\begin{equation}
\bm{j} = \frac{2\hbar en_s}{m^*}(\bm{\nabla} \phi - \frac{2e}{\hbar c}\bm{A}),
\label{eq:GL1}
\end{equation}
and the equation
\begin{equation}
\frac{1}{2m^*} (-i\hbar \bm{\nabla} -\frac{2e}{c}\bm{A})^2 \psi +\alpha(T) \psi +\beta(T) |\psi|^2\psi = 0.
\label{eq:GL2}
\end{equation}
The GL coherence length or magnetic length $\ell_c$ characterizes variation of the superconducting order parameter in the real space, and defined as 
\begin{equation}
\ell_c= \sqrt{\hbar^2/2m^*|\alpha|}.
\label{eq:coherence}
\end{equation}
The upper critical field $H_{c2}$ of type-II superconductor can be estimated from Eq.~(\ref{eq:GL2}). When the external applied magnetic field is close to $H_{c2}$, the superconducting order parameter $\psi$ becomes small, and Eq.~(\ref{eq:GL2}) can be linearized into 
\begin{equation}
\frac{1}{2m^*} (-i\hbar \bm{\nabla} -\frac{2e}{c}\bm{A})^2 \psi +\alpha \psi = 0,
\label{eq:LGL2}
\end{equation}
which resembles the Schr\"odinger equation for a particle with mass $m^*$ and charge $2e$ subject to magnetic field $\bm{B} = \nabla \times \bm{A}$. The solution of Eq.~(\ref{eq:LGL2}) gives rise to Landau levels of the corresponding particle \begin{equation}
|\alpha| = \hbar \omega_c (n+\frac{1}{2}),
\end{equation}
where $n$ are positive integers and $\omega_c = 2eB/m^*c$ is the cyclotron frequency. $H_{c2}$ is defined as the maximum magnetic field for the solution, namely, 
\begin{equation}
H_{c2} = m^*c|\alpha|/\hbar e = \frac{\Phi_s}{2\pi \ell_c^2},
\label{eq:Hc2_1}
\end{equation}
where $\Phi_s = \pi \hbar c/e \approx 2.067 \times 10^{-15}$ T$\cdot$m$^2$ is the magnetic (superconducting) quantum flux.

We show that $\ell_c$ and $H_{c2}$ can be obtained from the free energy for finite-momentum pairing superconducting state. For pairing wavevector $\bm{q}$, the real-space order parameter can be written as $
\psi = \sqrt{n_s(\bm{q})} e^{i\bm{q}\cdot \bm{r}}$. In the absence of external magnetic field, the superconducting condensation energy is reduced to
\begin{equation}
\delta F (\bm{q}) = F_s (\bm{q})  - F_n =  \frac{\hbar^2\bm{q}^2}{2m^*} n_s(\bm{q}) +\alpha n_s(\bm{q}) + \frac{1}{2} \beta n_s^2(\bm{q}),
\label{eq:feng1}
\end{equation}
and the relation of Eq.~(\ref{eq:GL2}) 
\begin{equation}
\frac{\hbar^2\bm{q}^2}{2m^*}  +\alpha+  \beta n_s(\bm{q}) =0.
\label{eq:GLr2}
\end{equation}
Therefore,  Eq.~(\ref{eq:feng1}) and Eq.~(\ref{eq:GLr2}) result in
\begin{equation}
\delta F (\bm{q}) = -\frac{1}{2\beta}  (|\alpha| - \frac{\hbar^2\bm{q}^2}{2m^*})^2 = \delta F(0) [1-(\bm{q}/q_c)^2]^2,
\label{eq:condeng}
\end{equation}
where $q_c = \sqrt{2m^*|\alpha|/\hbar^2}$ is the critical pairing wavevector defined by $\delta F (q_c) = 0$. Based on Eqs.~(\ref{eq:coherence}) and (\ref{eq:Hc2_1}), we have $ q_c =1/ \ell_c$, and 
\begin{equation}
H_{c2} = \frac{\Phi_s}{2\pi \ell_c^2} = \Phi_s q_c^2/2\pi.
\label{eq:Hc2_2}
\end{equation}
The supercurrent density of Eq.~(\ref{eq:GL1}) is 
\begin{equation}
\bm{j} = 2\hbar en_s(\bm{q}) \bm{q}/m^* = 2e n_s(\bm{q})\bm{v},
\end{equation}
where $\bm{v} = \hbar \bm{q}/m^*$ is the velocity of Cooper pair. For the present study, we can calculate the free energy as a function of pairing momentum within the microscopic model for MATBG superconductor. Based on Eq.~(\ref{eq:feng1}), 
\begin{equation}
\frac{\partial F(\bm{q})}{\partial (\hbar \bm{q}) }=  \frac{\hbar \bm{q}}{m^*} n_s(\bm{q}) + \left[\frac{\hbar^2\bm{q}^2}{2m^*}  +\alpha+  \beta n_s(\bm{q}) \right] \frac{\partial n_s(\bm{q})  }{\partial (\hbar \bm{q}) } =  n_s(\bm{q}) \bm{v},
\end{equation}
and we have the supercurrent density
\begin{equation}
\bm{j} = \frac{2e}{\hbar} \left[\frac{\partial F(\bm{q})}{\partial  \bm{q} }\right].
\label{eq:jc}
\end{equation}
In the above derivations, $\alpha$ and $\beta$ are assumed to be independent of pairing wavevector $\bm{q}$. Such a  assumption in MATBG can be justified by Fig.~1(c) in the main text, where our numerical calculations of $\delta F(\bm{q})$ can be well described by Eq.~(\ref{eq:condeng}).

\section{Interaction Hamiltonian}
\label{sec:InH}
We focus on electron pairing mainly mediated by in-plane optical phonons in graphene, which are thought to contribute $\sim70\%$ of the total electron-phonon coupling strength \cite{Choi:2018aa}. Based on $\mathcal{C}_{6v}$ and time-reversal symmetry, the in-plane optical phonons of isolated graphene can be classified into doubly degenerate $E_2$ modes at BZ center, and $A_1$, $B_1$ modes at BZ corner \cite{Basko:2008aa}. 
The optical phonon energies $\hbar \omega_{E_2} = 196$ meV and $\hbar \omega_{A_1} = \hbar \omega_{B_1} = 170$ meV are much higher than the kinetic energy of flat-band electrons in MATBG, indicating the optical phonon-mediated interaction is essentially instantaneous. By further restricting to electron pairing between opposite valley and spin, the optical phonon-mediated attraction 
\begin{equation}
\begin{aligned}
H_{ep} & = -2g_{0} \sum_{l \tau s \sigma }  \int d\bm{r}  \psi^{\dagger}_{l \tau s \sigma}(\bm{r}) \psi^{\dagger}_{l \bar{\tau} \bar{s}\sigma }(\bm{r}) \psi_{l \bar{\tau} \bar{s} \bar{\sigma} }(\bm{r}) \psi_{l \tau s \bar{\sigma} }(\bm{r}) -2g_{1} \sum_{l \tau s \sigma \sigma' }  \int d\bm{r}  \psi^{\dagger}_{l \tau s \sigma}(\bm{r}) \psi^{\dagger}_{l \bar{\tau} \bar{s} \sigma' } (\bm{r})  \psi_{l \tau \bar{s} \bar{\sigma}' }(\bm{r}) \psi_{l \bar{\tau} s \bar{\sigma}}(\bm{r}),
\end{aligned}
\end{equation}
where $\bar{\tau} = -\tau$, $\bar{s} = -s$, $\bar{\sigma} = -\sigma$, $g_{0}$ and $g_{1}$ are estimated to be 52 and 69 meV$\cdot$nm$^2$, giving the electron-electron interacting strengths mediated by $E_2$ and $A_1(B_1)$ phonons, respectively \cite{Wu:2018aa}. The real-space field operator can be expanded by plane-wave basis via Fourier transformation 
\begin{equation}
\psi_{l \tau s \sigma}(\bm{r}) = \frac{1}{\sqrt{A}} \sum_{\bm{k}} \psi_{l \tau s \sigma}(\bm{k})e^{i(\bm{k} +\bm{K}_{l \tau}) \cdot \bm{r}},
\end{equation} 
where $A$ is the area of the system, $\bm{K}_{l \tau}$ denotes the wavevector of the Dirac point labeled by $l$ and valley $\tau$. In the momentum space 
\begin{equation}
\begin{aligned}
H_{ep}  = &- 2 \sum_{\bm{k} \bm{k}' \bm{q}} \sum_{l \tau s \sigma } \left[ g_{0} \psi^{\dagger}_{l\tau s \sigma}(\bm{k}) \psi^{\dagger}_{l\bar{\tau} \bar{s}\sigma }(\bm{q}-\bm{k}) \psi_{l\bar{\tau} \bar{s} \bar{\sigma} }(\bm{q}-\bm{k}') \psi_{l\tau s \bar{\sigma} }(\bm{k}') \right. \\
&~~~~~~~~~~~~~~+ g_{1}  \sum_{\sigma' }  \left.  \psi^{\dagger}_{l \tau s \sigma}(\bm{k}) \psi^{\dagger}_{l \bar{\tau} \bar{s} \sigma' } (\bm{q}-\bm{k})  \psi_{l \tau \bar{s} \bar{\sigma}' }(\bm{q}-\bm{k}') \psi_{l \bar{\tau} s \bar{\sigma}}(\bm{k}') \right].
\end{aligned}
\label{eq:ephm}
\end{equation}
The Hamiltonian in the moir\'e momentum representation can be obtained by performing the following substitution 
\begin{equation}
\begin{aligned}
\sum_{\bm{k}} \psi_{l \tau s \sigma}(\bm{k}) &\rightarrow \sum_{n}  \sum_{\bm{k} \in  \text{MBZ}} \psi_{l \tau s \sigma n}(\bm{k} )
\end{aligned}
\end{equation} 
where $\bm{k}$ on the right hand side is measured from the $\gamma$ point of MBZ as depicted in Fig.~\ref{fig:figs1}, and 
\begin{equation}
\psi_{l \tau s \sigma n}(\bm{k} ) = \psi_{l \tau s \sigma}(\bm{k}-l\tau \bm{q}_1+\bm{G}_{n} ).
\end{equation}
In the moir\'e momentum representation, Eq.~(\ref{eq:ephm}) is reduced to
\begin{equation}
\begin{aligned}
H_{ep}  =&-2\sum_{l \tau \tau' s \sigma \sigma' } \sum_{nn'n_1n_1'} \sum_{\bm{q}\bm{k}\bm{k}' \in \text{MBZ}} (g_{0}\delta_{\tau \tau'} \delta_{\sigma \sigma'} +g_1 \delta_{\tau \bar{\tau}'} )\delta_{n+n_1,n'+n_1'} \psi^{\dagger}_{l \tau s \sigma n}(\bm{k}) \psi^{\dagger}_{l \bar{\tau} \bar{s} \sigma' n_1} (\bm{q}-\bm{k})  \psi_{l \bar{\tau}' \bar{s} \bar{\sigma}' n_1'}(\bm{q}-\bm{k}') \psi_{l \tau' s \bar{\sigma} n'}(\bm{k}') \\
=&-2\sum_{\alpha \alpha_1 \alpha'_1 \alpha'} g_{\alpha \alpha_1 \alpha'_1 \alpha'} \sum_{\bm{q}\bm{k}\bm{k}' \in \text{MBZ}}  \psi^{\dagger}_{\alpha}(\bm{k}) \psi^{\dagger}_{\alpha_1} (\bm{q}-\bm{k})  \psi_{\alpha_1'}(\bm{q}-\bm{k}') \psi_{\alpha'}(\bm{k}') ,
\end{aligned}
\label{eq:epcmr}
\end{equation}
where the collective label $\alpha =(l\tau s \sigma n)$, $\delta_{n+n_1,n'+n_1'}$ is imposed by the momentum conservation, and the interaction matrix element
\begin{equation}
g_{\alpha \alpha_1 \alpha'_1 \alpha'} = (g_{0}\delta_{\tau \tau'} \delta_{\sigma \sigma_1} +g_1 \delta_{\tau \bar{\tau}'} ) \delta_{ll'} \delta_{l_1l_1'} \delta_{ll_1} \delta_{\tau \bar{\tau}_1} \delta_{\tau' \bar{\tau}'_1} \delta_{ss'} \delta_{s_1s_1'} \delta_{s\bar{s}_1} \delta_{\sigma \bar{\sigma}'} \delta_{\sigma_1 \bar{\sigma}'_1} \delta_{n+n_1,n'+n_1'}.
\end{equation}
The model electron-electron interaction of Eq.~(2) in the main text becomes 
\begin{equation}
\begin{aligned}
H_{ee} & = 2u\sum_{ll'} \sum_{\tau s \sigma \sigma' } \int d\bm{r} \psi^{\dagger}_{l \tau s \sigma}(\bm{r}) \psi^{\dagger}_{l' \bar{\tau} \bar{s} \sigma' }(\bm{r})   \psi_{l' \bar{\tau} \bar{s} \sigma' }(\bm{r})  \psi_{l 
\tau s \sigma}(\bm{r}) \\
& = 2u\sum_{ll' \tau s \sigma \sigma' } \sum_{nn'n_1n_1'} \sum_{\bm{q}\bm{k}\bm{k}' \in \text{MBZ}}  \psi^{\dagger}_{l \tau s \sigma n}(\bm{k}) \psi^{\dagger}_{l' \bar{\tau} \bar{s} \sigma' n_1}(\bm{q}-\bm{k})   \psi_{l' \bar{\tau} \bar{s} \sigma' n_1'}(\bm{q}-\bm{k}')  \psi_{l \tau s \sigma n'}(\bm{k}') \\
& = 2\sum_{\alpha \alpha_1 \alpha'_1 \alpha'}  u_{\alpha \alpha_1 \alpha'_1 \alpha'}  \sum_{\bm{q}\bm{k}\bm{k}' \in \text{MBZ}} \psi^{\dagger}_{\alpha}(\bm{k}) \psi^{\dagger}_{\alpha_1} (\bm{q}-\bm{k})  \psi_{\alpha_1'}(\bm{q}-\bm{k}') \psi_{\alpha'}(\bm{k}') ,
\end{aligned}
\label{eq:eecmr}
\end{equation}
where $u$ is Coulomb repulsion strength, and the interacting matrix element 
\begin{equation}
u_{\alpha \alpha_1 \alpha'_1 \alpha'} = (u \delta_{\tau \tau'} )\delta_{ll'} \delta_{l_1l_1'} \delta_{\tau \bar{\tau}_1} \delta_{\tau' \bar{\tau}'_1} \delta_{ss'} \delta_{s_1s_1'} \delta_{s\bar{s}_1} \delta_{\sigma \sigma'} \delta_{\sigma_1 \sigma'_1} \delta_{n+n_1,n'+n_1'}.
\end{equation}
Based on Eqs.~(\ref{eq:epcmr}) and (\ref{eq:eecmr}), the total interacting Hamiltonian can be organized in a compact form as
\begin{equation}
\begin{aligned}
H_{in} &= H_{ep} +H_{ee} = \frac{1}{2}\sum_{\alpha \alpha_1 \alpha'_1 \alpha'}   \sum_{\bm{q}\bm{k}\bm{k}' \in \text{MBZ}} [V (\bm{k},\bm{k}')]_{\alpha \alpha_1 \alpha'_1 \alpha'} \psi^{\dagger}_{\alpha}(\bm{k}) \psi^{\dagger}_{\alpha_1} (\bm{q}-\bm{k})  \psi_{\alpha_1'}(\bm{q}-\bm{k}'), \psi_{\alpha'}(\bm{k}') 
\end{aligned}
\label{eq:tinmr}
\end{equation}
with $ [V (\bm{k},\bm{k}')]_{\alpha \alpha_1 \alpha'_1 \alpha'}  = 4 (u_{\alpha \alpha_1 \alpha'_1 \alpha'}- g_{\alpha \alpha_1 \alpha'_1 \alpha'})$ independent of momentum $\bm{k}$ and $\bm{k}'$.

\section{BdG Hamiltonian with finite-momentum pairing}
Within the mean-field theory, the total interacting Hamiltonian of Eq.~(\ref{eq:tinmr}) can be decoupled into finite-momentum pairing channel as
\begin{equation}
H_{in} = \frac{1}{2} \sum_{\alpha \alpha_1} \sum_{\bm{k} \in \text{MBZ}}  \left\{\psi^{\dagger}_{\alpha}(\bm{k}) \left[\Delta(\bm{q})\right]_{\alpha \alpha_1} \psi^{\dagger}_{\alpha_1} (\bm{q}-\bm{k}) + \text{H.c.} \right\}+ C_{\bm{q}},
\end{equation}
where $\bm{q}$ is the pairing momentum with the pair potential defined as
\begin{equation}
[\Delta_{\bm{q}}(\bm{k})]_{\alpha \alpha_1} =   \sum_{ \alpha'_1 \alpha'} \sum_{\bm{k}'}  [V(\bm{k},\bm{k}')]_{\alpha \alpha_1 \alpha'_1 \alpha'} \langle \psi_{\alpha_1'}(\bm{q}-\bm{k}') \psi_{\alpha'}(\bm{k}')  \rangle
\end{equation}
\begin{equation}
C_{\bm{q}}=  -\frac{1}{2}\sum_{\alpha \alpha_1 \alpha'_1 \alpha'} \sum_{\bm{k}\bm{k}' \in \text{MBZ}} [V(\bm{k},\bm{k}')]_{\alpha \alpha_1 \alpha'_1 \alpha'}  \langle \psi^{\dagger}_{\alpha}(\bm{k}) \psi^{\dagger}_{\alpha_1} (\bm{q}-\bm{k}) \rangle \langle   \psi_{\alpha_1'}(\bm{q}-\bm{k}') \psi_{\alpha'}(\bm{k}') \rangle. 
\end{equation}
Given the the anticommutation relation of fermionic operator, the matrix form of the superconducting gap satisfies the relation $\Delta_{\bm{q}}(\bm{k}) = -\Delta^{\text{T}}_{\bm{q}}(\bm{k})$. In the Nambu spinor representation, the total Hamiltonian
\begin{equation}
\begin{aligned}
H = & \frac{1}{2}\sum_{\bm{k}\in \text{MBZ}} \Psi_{\bm{q}}^{\dagger}(\bm{k}) \mathcal{H}_{\text{BdG}}(\bm{q},\bm{k}) \Psi_{\bm{q}}(\bm{k}) + \frac{1}{2} \sum_{\bm{k}\in \text{MBZ}} \text{Tr}\left[ \mathcal{H}(\bm{q}-\bm{k})-\mu \right] +C_{\bm{q}},
\end{aligned}
\end{equation}
where $\Psi_{\bm{q}}(\bm{k}) = \left[ \psi(\bm{k}), \psi^{\dagger}(\bm{q}-\bm{k}) \right]^{\text{T}}$, and the BdG Hamiltonian
\begin{equation}
\mathcal{H}_{\text{BdG}}(\bm{q},\bm{k})  =
\begin{bmatrix}
\mathcal{H}(\bm{k})-\mu   & \Delta_{\bm{q}}(\bm{k}) \\
\Delta^{\dagger}_{\bm{q}}(\bm{k}) & -\mathcal{H}^{\text{T}}(\bm{q}-\bm{k})-\mu 
\end{bmatrix}.
\end{equation}

The self-consistent gap equation can be formulated by using the Green's function defined as \cite{Mahan:2013aa}
\begin{equation}
\begin{aligned}
G(\tau,\bm{q},\bm{k}) &= -\langle T_{\tau} \Psi_{\bm{q}}(\tau,\bm{k}) \Psi_{\bm{q}}^{\dagger}(0,\bm{k}) \rangle  =
 \begin{bmatrix}
 \mathcal{G}(\tau,\bm{k}) & \mathcal{F}_{\bm{q}}(\tau,\bm{k}) \\
 \mathcal{F}_{\bm{q}}^{\dagger}(-\tau,\bm{k}) & - \mathcal{G}^{\text{T}}(-\tau,\bm{q}-\bm{k})
 \end{bmatrix}
\end{aligned}
\end{equation} 
where $\tau$ denotes imaginary time, $T_{\tau}$ is the time-order operator, $\mathcal{G}(\tau,\bm{k}) =  -\langle T_{\tau} \psi(\tau,\bm{k}) \psi^{\dagger}(0,\bm{k}) \rangle$  and $ \mathcal{F}_{\bm{q}}(\tau,\bm{k}) = -\langle T_{\tau} \psi(\tau,\bm{k}) \psi(0,\bm{q}-\bm{k}) \rangle$ are the normal and Gorkov's anomalous Green's functions, respectively.
By performing Fourier transformation on $\tau$, we have
\begin{equation}
\begin{aligned}
G(\tau,\bm{q},\bm{k}) = k_B T  \sum_{n} e^{-i\omega_n \tau} \left[i\omega_n - \mathcal{H}_{\text{BdG}}(\bm{q},\bm{k})\right]^{-1} = k_B T  \sum_{n} e^{-i\omega_n \tau} U_{\bm{q}}(\bm{k}) \left[i\omega_n -E_{\bm{q}}(\bm{k})\right]^{-1} U_{\bm{q}}^{\dagger}(\bm{k}),
\end{aligned}
\end{equation}
wehre $k_B$ is the Boltzmann constant, $T$ denotes temperature, $\omega_{n} = (2n+1)\pi k_B T$ with integer $n$, and $U_{\bm{q}}(\bm{k})$ is the unitary matrix that diagonalizes $\mathcal{H}_{\text{BdG}}(\bm{q},\bm{k})$, giving the quasiparticle energy spectrum $E_{\bm{q}}(\bm{k})$. Therefore, we have

\begin{equation}
\begin{aligned}
G(0,\bm{q},\bm{k}) &=  U_{\bm{q}}(\bm{k}) k_B T \sum_{n}  \left[i\omega_n -E_{\bm{q}}(\bm{k})\right]^{-1} U_{\bm{q}}^{\dagger}(\bm{k}).
\end{aligned}
\end{equation}
Hereafter and in the main text, we use $G(\bm{q},\bm{k}) $ to denote $G(0,\bm{q},\bm{k}) $ and have
\begin{equation}
G(\bm{q},\bm{k})  = 
\begin{bmatrix}
\mathcal{G}(\bm{k}) & \mathcal{F}_{\bm{q}}(\bm{k}) \\
\mathcal{F}_{\bm{q}}^{\dagger}(\bm{k})  & -\mathcal{G}^{\text{T}}(\bm{q} - \bm{k}) 
\end{bmatrix} = U_{\bm{q}}(\bm{k}) f[E_{\bm{q}}(\bm{k})] U_{\bm{q}}^{\dagger}(\bm{k}),
\end{equation}
where $f(\epsilon)$ is the Fermi-Dirac distribution function. Obviously, the normal Green's function also depends on the pairing momentum $\bm{q}$, and the self-consistent gap equation is
\begin{equation}
[\Delta_{\bm{q}}(\bm{k})]_{\alpha \alpha_1} = \sum_{\alpha' \alpha_1'}  \sum_{\bm{k} \in \text{MBZ}} [V(\bm{k},\bm{k}')]_{\alpha  \alpha_1\alpha_1' \alpha'} [\mathcal{F}_{\bm{q}}(\bm{k})]_{\alpha' \alpha_1'}, 
\end{equation}
which is independent of $\bm{k}$ for the present model study. Hereafter, we simply use $\Delta_{\bm{q}}$ to denote the pair potential.
\begin{equation}
C_{\bm{q}} =  - \frac{1}{2}\sum_{\alpha \alpha_1 \alpha'_1 \alpha'}  \sum_{\bm{k}\bm{k}' \in \text{MBZ}} [V(\bm{k},\bm{k}')]_{\alpha  \alpha_1\alpha_1' \alpha'} [\mathcal{F}_{\bm{q}}^{\dagger}(\bm{k})]_{\alpha_1 \alpha} [\mathcal{F}_{\bm{q}}(\bm{k}')]_{\alpha' \alpha_1'} =- \frac{1}{2} \text{Tr}\mathcal{F}^{\dagger}_{\bm{q}} V \mathcal{F}_{\bm{q}} ,
\end{equation}
where $V$ is interacting tensor matrix. To insist the particle number conservation, the chemical potential for finite-momentum pairing state should be determined self consistently by
\begin{equation}
N = \sum_{\bm{k} \in \text{MBZ} } \text{Tr}[ \mathcal{G}(\bm{k})],
\end{equation}
with $N$ the total number of the particles in the system.

Although the above lumped label $\alpha = (l\tau s \sigma n)$ is convenient for formulating, the detailed expression for pair potential $\Delta_{\bm{q}}$ helps understanding the corresponding physical means. For example, 
\begin{equation}
\begin{aligned}
~[\Delta_{\bm{q}}]_{l\tau s \sigma n, l_1\bar{\tau}\bar{s}\sigma_1n_1} &=   4\sum_{\tau' \sigma' \sigma_1'} \sum_{n' n_1'} \sum_{\bm{k}'} [u \delta_{\tau \tau'} \delta_{\sigma \sigma'} \delta_{\sigma_1 \sigma'_1}  - (g_{0}  \delta_{\tau \tau'} \delta_{\sigma \sigma_1} + g_1\delta_{\tau \bar{\tau}'}) \delta_{ll_1} \delta_{\sigma \bar{\sigma}'} \delta_{\sigma_1 \bar{\sigma}'_1}]\delta_{n+n_1,n'+n_1'} \\
& \times  \langle \psi_{l_1\bar{\tau}'\bar{s} \sigma_1' }(\bm{q}-\bm{k}'+l_1\tau'\bm{q}_1+\bm{G}_{n_1'}) \psi_{l \tau' s \sigma' }(\bm{k}'-l\tau'\bm{q}_1+\bm{G}_{n'})  \rangle.
\end{aligned}
\end{equation}
Due to the momentum conservation, we can use a single reciprocal lattice vector $\bm{Q} = \bm{G}_n + \bm{G}_{n_1}$ to label $\Delta_{\bm{q}}$ instead of two of $(\bm{G}_n, \bm{G}_{n_1})$, namely,
\begin{equation}
\begin{aligned}
~[\Delta_{\bm{q},\bm{Q}} ]_{l\tau s \sigma, l_1\bar{\tau}\bar{s}\sigma_1} &=   4\sum_{\tau' \sigma' \sigma_1'} \sum_{n' \bm{k}'} [u \delta_{\tau \tau'} \delta_{\sigma \sigma'} \delta_{\sigma_1 \sigma'_1}  - (g_{0}  \delta_{\tau \tau'} \delta_{\sigma \sigma_1} + g_1\delta_{\tau \bar{\tau}'}) \delta_{ll_1} \delta_{\sigma \bar{\sigma}'} \delta_{\sigma_1 \bar{\sigma}'_1}] \\
& \times  \langle \psi_{l_1\bar{\tau}'\bar{s} \sigma_1' }(\bm{q}-\bm{k}'+l_1\tau'\bm{q}_1+\bm{Q}- \bm{G}_{n'}) \psi_{l \tau' s \sigma' }(\bm{k}'-l\tau'\bm{q}_1+\bm{G}_{n'})  \rangle.
\end{aligned}
\end{equation}
Therefore the intralayer pair potential 
\begin{equation}
\begin{aligned}
~[\Delta_{\bm{q},\bm{Q}} ]_{l\tau s \sigma, l\bar{\tau}\bar{s}\sigma_1} &=   4u  \sum_{n' \bm{k}'}  \langle \psi_{l \bar{\tau} \bar{s} \sigma_1}(\bm{q}-\bm{k}'+l \tau \bm{q}_1+\bm{Q}- \bm{G}_{n'}) \psi_{l \tau s \sigma }(\bm{k}'-l\tau \bm{q}_1+\bm{G}_{n'})  \rangle \\
 &- 4g_{0}  \sum_{n' \bm{k}'} \langle \psi_{l \bar{\tau} \bar{s} \bar{\sigma}_1 }(\bm{q}-\bm{k}'+l\tau \bm{q}_1+\bm{Q}- \bm{G}_{n'}) \psi_{l \tau s \bar{\sigma} }(\bm{k}'-l\tau \bm{q}_1+\bm{G}_{n'})  \rangle \delta_{\sigma \sigma_1} \\
 &- 4g_1\sum_{n' \bm{k}'}   \langle \psi_{l \tau \bar{s} \bar{\sigma}_1 }(\bm{q}-\bm{k}'-l\tau \bm{q}_1+\bm{Q}- \bm{G}_{n'}) \psi_{l \bar{\tau} s \bar{\sigma} }(\bm{k}'+l\tau \bm{q}_1+\bm{G}_{n'})  \rangle.
\end{aligned}
\label{eq:intralayergap}
\end{equation}
The interlayer pair potential
\begin{equation}
\begin{aligned}
~[\Delta_{\bm{q},\bm{Q}} ]_{l\tau s \sigma, \bar{l}\bar{\tau}\bar{s}\sigma_1} &=   4u  \sum_{n' \bm{k}'} \langle \psi_{\bar{l}\bar{\tau} \bar{s} \sigma_1 }(\bm{q}-\bm{k}'-l \tau \bm{q}_1+\bm{Q}- \bm{G}_{n'}) \psi_{l \tau s \sigma }(\bm{k}'-l \tau \bm{q}_1+\bm{G}_{n'})  \rangle.
\end{aligned}
\label{eq:interlayergap}
\end{equation}
In the sublattice space, the matrix form of pairing potential
\begin{equation}
\Delta_{\bm{q},\bm{Q}} =
\begin{pmatrix}
[\Delta_{\bm{q},\bm{Q}}]_{\text{AA}} & [\Delta_{\bm{q},\bm{Q}}]_{\text{AB}} \\
[\Delta_{\bm{q},\bm{Q}}]_{\text{BA}} & [\Delta_{\bm{q},\bm{Q}}]_{\text{BB}}
\end{pmatrix}.
\end{equation}
In the present study, numerical calculations are carried out by taking a cutoff on the reciprocal lattice vector $\bm{Q}$ since the magnitude of pair potential decreases quickly upon increasing $\bm{Q}$. In Fig.~\ref{fig:figs1}, we depict representative reciprocal lattice vectors $\bm{Q}_{1,2,3,4}$. All of the reciprocal lattice vectors generated by $\mathcal{P} \bm{Q}_{n=1,2,3,4}$ are included in our calculations, where $\mathcal{P}$ are symmetry operations from point group $\mathcal{C}_6$.
\begin{figure}
\centering
\includegraphics[width=\columnwidth]{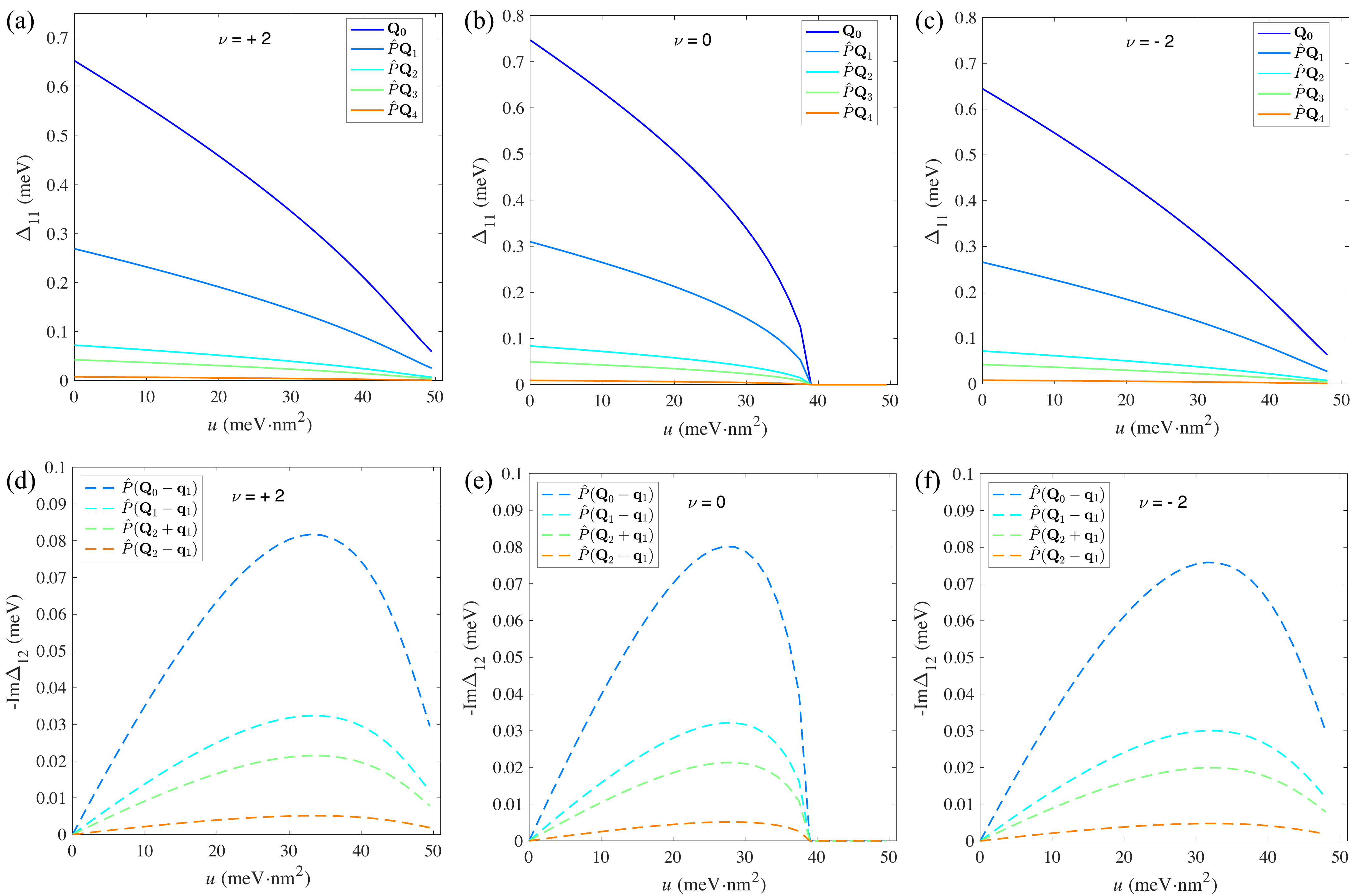}
\caption{Zero-momentum pair potential {\it vs.} the strength of electron-electron repulsion $u$ for MATBG with $\theta = 1.07^{\circ}$ and $\eta = 0.85$. The upper (lower) row show the intralayer (interlayer) intrasublattice pair potentials for different band fillings.}
\label{fig:figs2}
\end{figure}

\section{Pair potential and critical temperature}

Figure~\ref{fig:figs2} (a)-(c) show that the intralayer pair potential $\Delta_{11}$ is suppressed by increasing the electron-electron repulsion $u$. As suggested by Eq.~(\ref{eq:interlayergap}), the interlayer pair potential $\Delta_{12}$ can be labeled by vectors $\mathcal{P}(\bm{Q}-2l\tau\bm{q}_1)$. Figure.~\ref{fig:figs2} (d)-(f) show  nonmonotonous behaviors of $\Delta_{12}$ upon increasing $u$, attributed to the combined effects between optical phonon-mediated attraction and electron-electron repulsion. In addition to smaller magnitude, $\Delta_{12}$ also exhibits a $\pi/2$ phase difference comparing with $\Delta_{11}$.

\begin{figure}
\centering
\includegraphics[width=0.7\columnwidth]{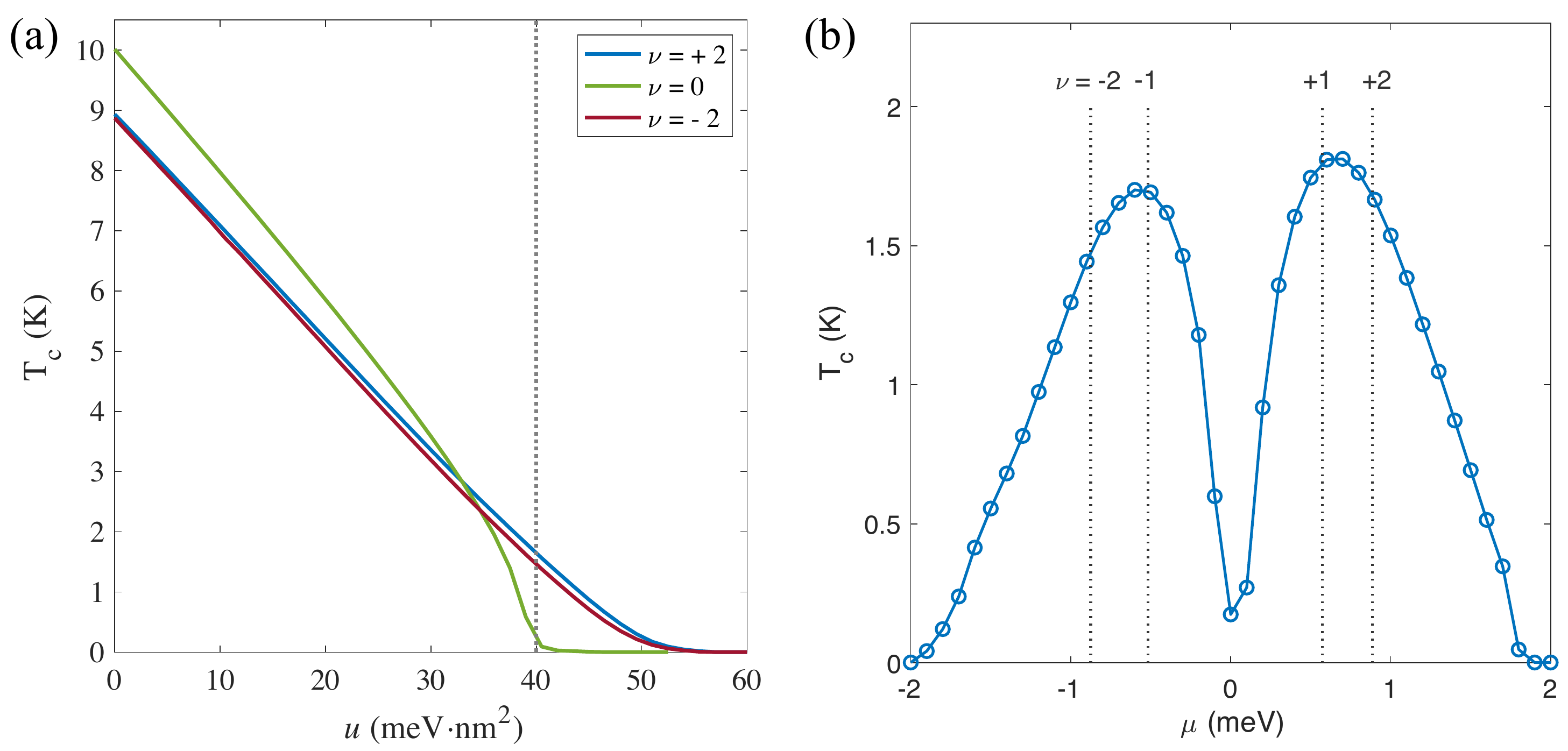}
\caption{Mean-field results of the critical temperature $T_c$ for MATBG with $\theta = 1.07^{\circ}$ and $\eta = 0.85$. (a) $T_c$ {\it vs.} $u$ for different band fillings. (b) $T_c$ {\it vs.} chemical potential $\mu$ for $u=40$ meV$\cdot$nm$^2$. }
\label{fig:figs3}
\end{figure}

The mean-field critical temperature $T_c$ is obtained identify the temperature where the pair potential vanishes. In Figure~\ref{fig:figs3} shows a typical results of $T_c$ as functions of $u$ and chemical potential $\mu$ or band filling $\nu$.

\begin{figure}
\centering
\includegraphics[width=0.75\columnwidth]{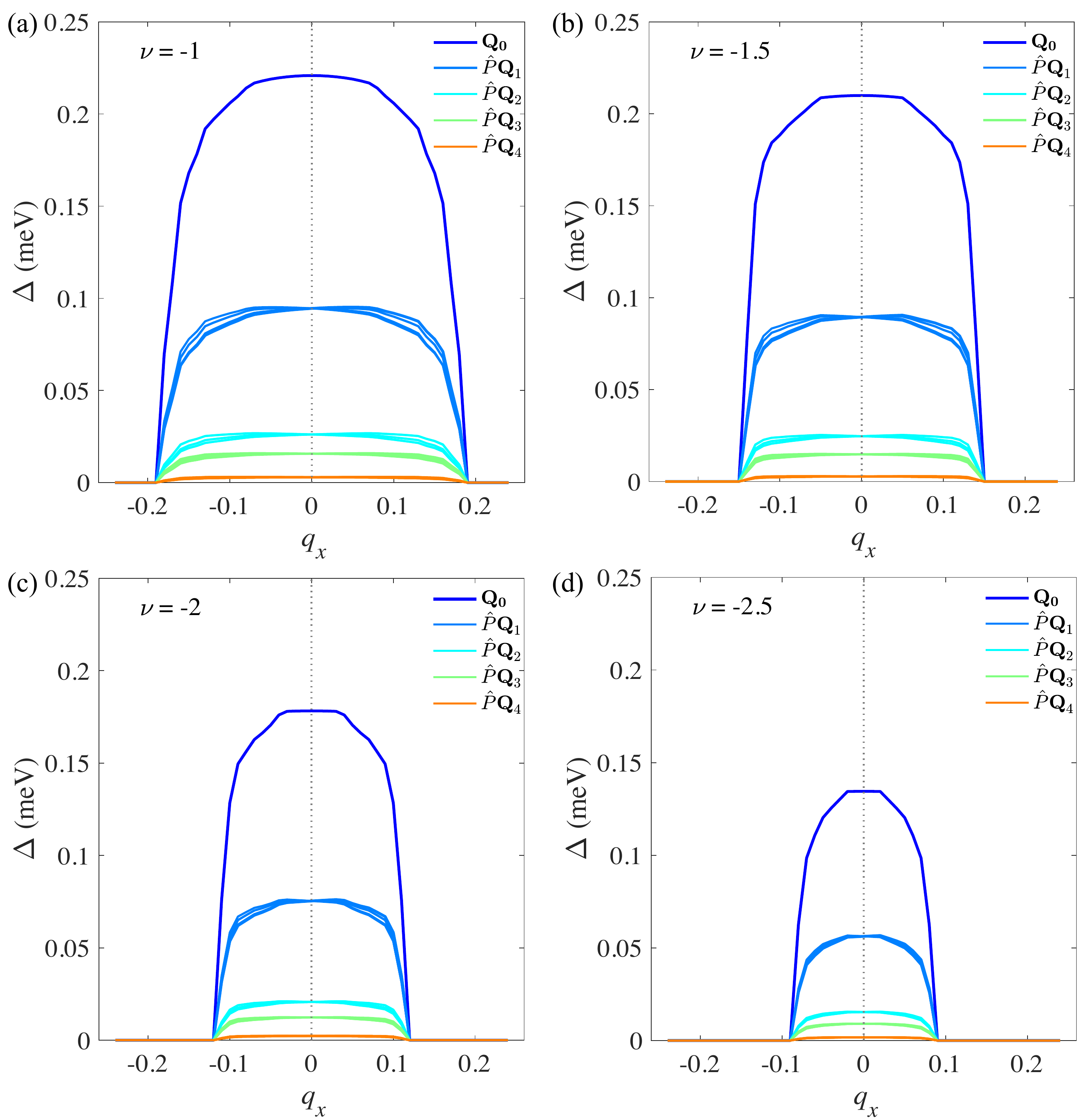}
\caption{(a)-(d) Intralayer intrasublattice pair potentials {\it vs.} pairing wavevector $q_x$ for MATBG with $\theta = 1.07^{\circ}$, $\eta = 0.85$ and band fillings (a) $\nu = -1$, (b) $\nu = -1.5$, (c) $\nu = -2$, (d) $\nu = -2.5$.}
\label{fig:fig4}
\end{figure}

As complementary results to Fig.~1(b) in the main text, Fig.~\ref{fig:fig4} plot pair potentiala as functions of paring wavevector for different band fillings. 

\section{Free energy of the superconducting state}

\begin{figure}
\centering
\includegraphics[width=0.375\columnwidth]{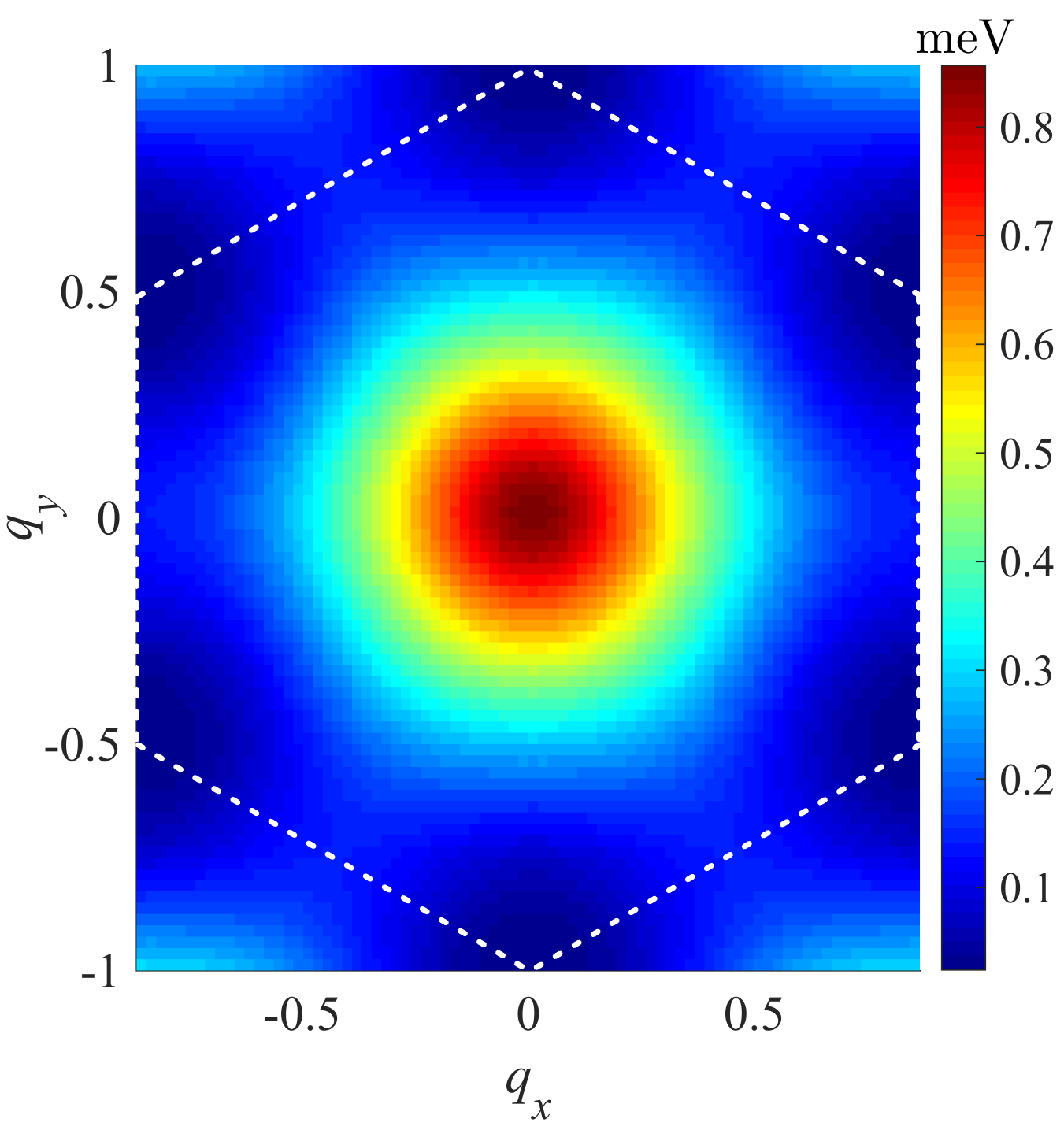}
\caption{Superconducting state condensation energy per moir\'e supercell {\it vs.} two-dimensional pairing wavevector $\bm{q}$ for MATBG with $\theta = 1.07^{\circ}$, $\eta = 0.85$, $u=0$ meV$\cdot$nm$^2$, and band filling $\nu = -1$. The MBZ is highlighted by dashed hexagon.}
\label{fig:figs5}
\end{figure}

The free energy of a superconducting system is defined microscopically  as
\begin{equation}
F_s(\bm{q},T) = \Omega[\Delta(\bm{q}),\mu_{\bm{q}},\bm{q},T]+N\mu_{\bm{q}},
\end{equation}
where the thermodynamic grand potential
\begin{equation}
\begin{aligned}
\Omega &= -\frac{1}{\beta} \ln{\text{Tr}(e^{-\beta H})},
\end{aligned}
\end{equation}
with $\beta = 1/k_BT$. For MATBG superconductor 
\begin{equation}
\begin{aligned}
F_s(\bm{q},T) & = -\frac{1}{\beta} \ln{\text{Tr} \left[e^{-\frac{\beta}{2}\sum_{\bm{k}\in \text{MBZ}} \Psi_{\bm{q}}^{\dagger}(\bm{k}) \mathcal{H}_{\text{BdG}}(\bm{q},\bm{k}) \Psi_{\bm{q}}(\bm{k}) + \frac{1}{2} \sum_{\bm{k}\in \text{MBZ}} \text{Tr}\left[ \mathcal{H}(\bm{q}-\bm{k})-\mu(\bm{q}) \right] +C_{\bm{q}}} \right] } + N \mu_{\bm{q}} \\
& = C_{\bm{q}} + N \mu_{\bm{q}} + \frac{1}{2} \sum_{\bm{k}\in \text{MBZ}} \text{Tr}\left[ \mathcal{H}(\bm{q}-\bm{k})-\mu_{\bm{q}} \right] - \frac{1}{2\beta} \sum_{\bm{k} \in \text{MBZ}} \text{Tr} \left\{ \ln \left[1+e^{-\beta E_{\bm{q}}(\bm{k})} \right] \right\} \\
& = C_{\bm{q}}  +A n_0 \mu_{\bm{q}} - \frac{1}{2\beta} \sum_{\bm{k} \in \text{MBZ}} \text{Tr} \left\{ \ln \left[1+e^{-\beta E_{\bm{q}}(\bm{k})} \right] \right\} \\
& =  C_{\bm{q}}  +A n_0 \mu_{\bm{q}}  + \frac{1}{2\beta} \text{Tr}  \sum_{\bm{k} \in \text{MBZ}} \ln f[- E_{\bm{q}}(\bm{k})], 
\end{aligned}
\end{equation}
where $n_0$ is the carrier density measured from the charge neutrality of MATBG, $A$ is area of the system. We have used the relation $\text{Tr}[ \mathcal{H}(\bm{q}-\bm{k})] = 0 $ since $\mathcal{H}(\bm{k})$ consists of massless Dirac Hamiltonian. The free energy density
\begin{equation}
f_s(\bm{q},T) = F_s(\bm{q},T)/A = C_{\bm{q}}/A + n_0 \mu_{\bm{q}}+ \frac{1}{8 \pi^2\beta}  \text{Tr} \int_{\text{MBZ} } d\bm{k} \ln f[- E_{\bm{q}}(\bm{k})].
\end{equation}
Figure.~\ref{fig:figs5} plot the superconducting condensation energy $\delta F(\bm{q})$, defined in the main text, for a special case of vanished electron-electron repulsion.

\section{Critical supercurrent}
\begin{figure}
  \centering
 \includegraphics[width= 0.6 \columnwidth ]{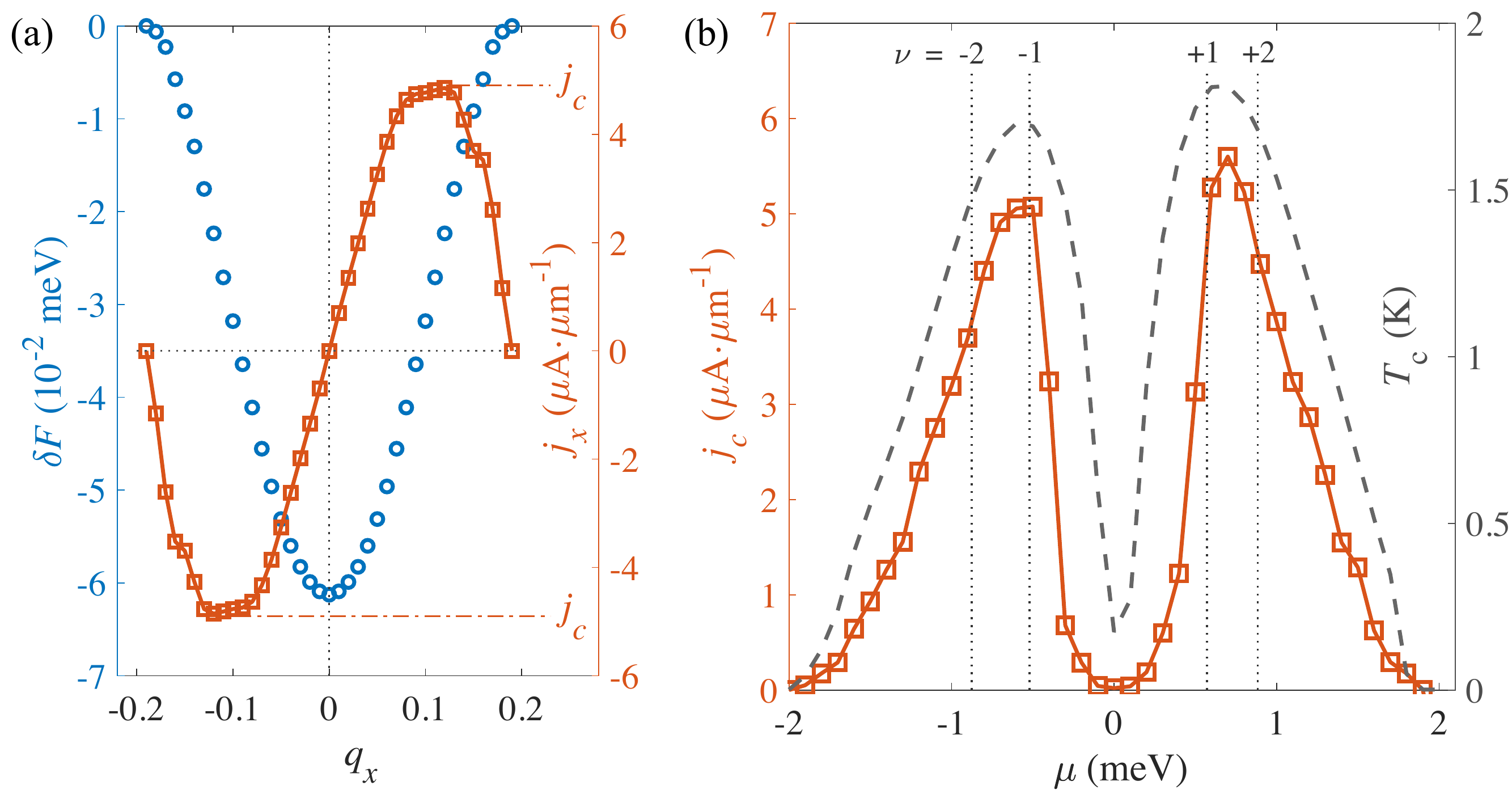}
  \caption{Supercurrent density for MATBG with $\theta = 1.07^{\circ}$, $\eta = 0.85$, and $u=40$ meV$\cdot$nm$^2$. (a) Superconducting state condensation energy (circles) and supercurrent density (squares) {\it vs.} pairing wavevector $q_x$ for $\nu = -1$. The horizontal dotted lines highlight the critical supercurrent density $j_c$. (b) $j_c$ {\it vs.} chemical potential $\mu$.The dashed curve gives the corresponding critical temperature $T_c$.} 
  \label{fig:figs6}
\end{figure}

Figure~\ref{fig:figs6}(a) shows a typical result of supercurrent density $j_x$ for MATBG calculated using Eq.~(\ref{eq:jc}). Here $j_x$ is an odd function of pairing wavevector $q_x$ due to time-reversal symmetry. As illustrated in Fig.~1(d) of the main text, $\delta F(\bm{q})$ is nearly isotropic, suggesting the critical supercurrent density $j_c$ can be defined by the maxima of $j_x$ as depicted in Fig.~\ref{fig:figs6}(a). Figure~\ref{fig:figs6}(b) plots $j_c$ and mean-field critical temperature $T_c$ as functions of chemical potential $\mu$. Similar to $T_c$, $j_c$ exhibits a dome-like feature. 

Table \ref{tab:tables1} summarizes experimental data on critical temperature $T_c$ and critical suppercurrent $I_c$. The critical suppercurrent density can be extracted by $j_c = I_c /\sqrt{A_c}$, where the conducting area $A_c$ of MATBG sample is $\sim$1 $\mu$m$^2$ \cite{Cao:2018aa,Yankowitz:2019aa,Lu:2019aa,Stepanov:2020aa,Saito:2020aa}. $I_c$ changes dramatically for MATBG with different twist angles and band fillings. For a given sample, e.g. $\theta = 1.1^{\circ}$ in Table \ref{tab:tables1}, a higher value of $T_c$ is associated with a larger value of $I_c$, being qualitatively consistent with our calculation shown in Fig.~\ref{fig:figs6}(b).  The quantitative value of $j_c$ in our calculations is over an order of magnitude larger than experimental results. There are two possible reasons that lead to suppressions on the experimentally measured $j_c$ for MATBG. Firstly, the strongest superconducting dome usually occurs in the band filling regime where two flavors are partially occupied associated with reconstructed flat bands \cite{Cao:2018aa,Yankowitz:2019aa,Lu:2019aa,Xie:2020ab}. Our model calculation does not include such electronic correlation-induced effect, resulting in an overestimated carrier density and supercurrent density. Secondly, the disorders in experimental system is inevitable. For MATBG superconductor, the presence of local charge disorder and twist angle disorder, as observed in a recent experimental work \cite{Uri:2020aa}, leads to the coexistence of superconducting and non superconducting regions, which form Josephson links and limit the zero-magnetic-field critical supercurrent. An experimental evidence for the Josephson links is the Fraunhofer-like quantum interference \cite{Cao:2018aa,Yankowitz:2019aa,Lu:2019aa,Stepanov:2020aa,Saito:2020aa}.

\begin{table}
\caption{Experimental results for the superconducting transition temperature $T_c$, and critical current $I_c$ for MATBG with different twist angle $\theta$, band filling $\nu$, and screening environment.}
\label{tab:tables1} 
\begin{ruledtabular}
\begin{tabular}{l  l l l l}
     $\theta$ ($^\circ $)  &  $\nu$ & $T_c$ (K) & $I_c$ (nA) & Refs \\
     \hline
     1.05  &  -1.82 & 1.2 &50  &\cite{Cao:2018aa}\\
     1.16  &  -2.15 & 0.5  & 17 &\\
     \hline
     1.27 (1.33 Gpa) & -2.33 & 3.1 & 120  &\cite{Yankowitz:2019aa}\\
     1.27 (2.21 Gpa) & -2.62 & 2 & 200  & \\
     
     \hline
           & -2.31 &  3.1 &  70   &\cite{Lu:2019aa} \\
  1.1     & -1  &  0.14 &  5  &\\
              & 0.67 &  0.16 & 7  &\\
             & 1.48 &  0.65 & 16  &\\
     \hline
      1.15   & -1.6  &  0.92 &300 &\cite{Stepanov:2020aa} \\  
                & 1.8  &  0.42  & 12  &\\
     \hline
     1.04   & -2.43  &  1.3 &  20 &\cite{Saito:2020aa} \\
     1.08   & -2.46  &  2.4 &  80  &\\
     1.09   & -2.79  &  2.5 & 20  &\\
     1.18   & -2.5  &  0.7 &  40 &\\
     1.12   & -2.47  &  4 & 160  &\\
\end{tabular}
\end{ruledtabular}
\end{table}

\section{Electron-acoustic phonon interaction}
The Fermi surface averaged electron-phonon spectrum function is defined as
 \begin{equation}
 \alpha^2F(\omega) \equiv  \frac{1}{N_{\mu}}\sum_{nm,\bm{k}\bm{k'}} |g_{nm}(\bm{k},\bm{k}')|^2 \delta(\epsilon_{n\bm{k}}-\mu)\delta(\epsilon_{m\bm{k}'}-\mu)\delta(\omega-\omega_{ph}), 
 \end{equation}
 where $m$ and $n$ are flat-band indices. Here $N_{\mu}$ is the density of states at the Fermi level $\mu$, $g_{nm}(\bm{k},\bm{k}')$ is flat-band projected electron-acoustic phonon interacting matrix, $\epsilon_{n\bm{k}}$ denotes electron energy for flat band $n$ and moir\'e momentum $\bm{k}$, and the phonon energy $\omega_{ph} =v_{ph} |\bm{k}-\bm{k}'|$ with $v_{ph} = 2\times10^4$ m/s the sound velocity for isolated graphene sheet. In the present study, $g_{nm}(\bm{k},\bm{k}')$ is obtained by folding the in-plane acoustic phonon mode of isolated graphene into MBZ, where the interlayer potential is ignored. By explicitly considering the interlayer potential, the two in-plane acoustic phonon modes from the two layers of graphene recombine into layer symmetric and asymmetric modes \cite{Koshino:2019aa}. We deliver the detailed derivation and expression of $g_{nm}(\bm{k},\bm{k}')$ in a separate work.

\begin{figure}
  \centering
 \includegraphics[width= 0.4 \columnwidth ]{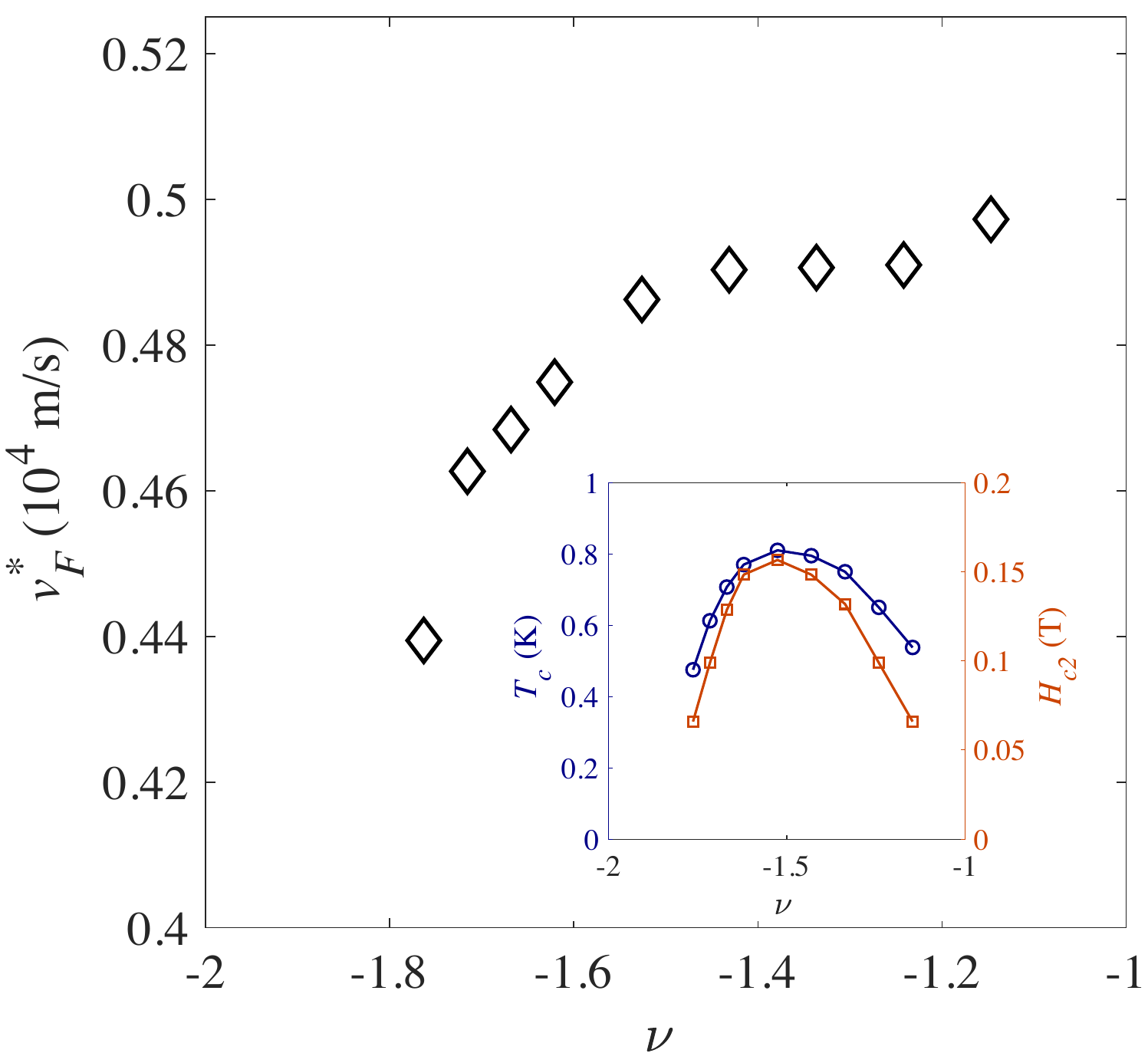}
  \caption{ Extracted $v_F^*$ from experimentally reported data of $T_c$ and perpendicular $H_{c2}$ shown in the insert, where $\theta = 1.15^{\circ}$ and electronic correlation is weaken by metallic screening \cite{Stepanov:2020aa}.} 
  \label{fig:figs7}
\end{figure}

\section{Experimental data}
Here we show a concrete example of extracting the average Fermi velocity from a recent experimental work on MATBG superconductor. As depicted in Fig.~\ref{fig:figs7}, $v_F^*$ does not show a negative correlation with $T_c$, which is usually expected for weak coupling superconductor with a given pairing glue. Therefore, we argue that the conventional weak pairing mechanism may be insufficient for explaining the exotic superconductivity observed in MATBG.

\section{Pairing channel}
In this section, we discuss the potential pairing channels for phonon-mediated superconductivity in MATBG. As discussed in Sec.~\ref{sec:InH}, the in-plane optical phonon-mediated electron-electron attraction
\begin{equation}
\begin{aligned}
H_{op} & = -2g_{0} \sum_{l \tau s s' \sigma }  \int d\bm{r}  \psi^{\dagger}_{l \tau s \sigma}(\bm{r}) \psi^{\dagger}_{l \bar{\tau} s'\sigma }(\bm{r}) \psi_{l \bar{\tau} s' \bar{\sigma} }(\bm{r}) \psi_{l \tau s \bar{\sigma} }(\bm{r}) -2g_{1} \sum_{l \tau ss' \sigma \sigma' }  \int d\bm{r}  \psi^{\dagger}_{l \tau s \sigma}(\bm{r}) \psi^{\dagger}_{l \bar{\tau} s' \sigma' } (\bm{r})  \psi_{l \tau s' \bar{\sigma}' }(\bm{r}) \psi_{l \bar{\tau} s \bar{\sigma}}(\bm{r}),
\end{aligned}
\end{equation}
 For pairing symmetry analysis, we ignore the retarded effect of acoustic phonon-mediated electron-electron attraction, which can then be described by
\begin{equation}
H_{ap} = -2g_a \sum_{l \tau s s' \sigma } \int d\bm{r} \psi^{\dagger}_{l \tau s \sigma}(\bm{r}) \psi^{\dagger}_{l \bar{\tau} s' \sigma'}(\bm{r})\psi_{l \bar{\tau} s' \sigma'}(\bm{r})\psi_{l \tau s \sigma}(\bm{r}), 
\end{equation} 
where $g_a$ represents the interaction strength. The electron-electron repulsion 
\begin{equation}
H_{ee} = 2u\sum_{ll'} \sum_{\tau s \sigma \sigma' } \int d\bm{r} \psi^{\dagger}_{l \tau s \sigma}(\bm{r}) \psi^{\dagger}_{l' \bar{\tau} s' \sigma' }(\bm{r})   \psi_{l' \bar{\tau} s' \sigma' }(\bm{r})  \psi_{l \tau s \sigma}(\bm{r}). 
\end{equation}
In contrast to Sec.~\ref{sec:InH}, here we release the restriction of electron pairing with opposite spins while still keep the assumption that electron pairing with opposite valleys. Within mean-field theory, the gap equation for real-space intrasublattice pair potential
\begin{equation}
\Delta_{+\sigma s,-\sigma s'} = g_0\langle \psi_{+\bar{\sigma} s}\psi_{-\bar{\sigma}s'}\rangle- g_1\langle \psi_{+\bar{\sigma}s'}\psi_{-\bar{\sigma}s}\rangle +(g_a-u)\langle \psi_{+\sigma s}\psi_{-\sigma s'}\rangle.
\end{equation}
The gap equation for intersublattice pair potential 
\begin{equation}
\Delta_{+\sigma s,-\bar{\sigma} s'} = - g_1\langle \psi_{+\sigma s'} \psi_{-\bar{\sigma}s}\rangle +(g_a-u)\langle \psi_{+\sigma s}\psi_{-\sigma s'}\rangle,
\end{equation}
indicating that the BZ center $E_2$ optical phonon modes do not contribute to intersublattice pairing. Table~\ref{tab:tables2} lists the pairing strengths for different intrasublattice pairing channels. Based on these results, we conclude that, for optical phonon-mediated superconductivity, the spin-singlet intrasublattice pairing is the dominant electron pairing channel. In the main text, we restrict ourselves to consider this dominant electron pairing channel. 

\begin{table}
\caption{Pairing strengths for different intrasublattice pairing channels.}
\label{tab:tables2} 
\begin{ruledtabular}
\begin{tabular}{c | c c c c}
   Pairing channel      & Valley triplet $\&$ Spin singlet  &  Valley singlet $\&$ Spin triplet & Equal spin \\

& $\langle \psi_{+\sigma \downarrow}\psi_{- \sigma \uparrow}\rangle = -\langle \psi_{+\sigma \uparrow}\psi_{- \sigma \downarrow} \rangle$  &  $\langle \psi_{+\sigma \downarrow}\psi_{-\sigma \uparrow}\rangle = \langle \psi_{+\sigma \uparrow}\psi_{-\sigma \downarrow} \rangle$& $\langle \psi_{+\sigma \downarrow}\psi_{-\sigma\downarrow}\rangle $   \\
    \hline
  Pairing strength & $g_0+g_1+g_a-u$  &  $g_0-g_1+g_a-u$ & $g_0-g_1+g_a-u$   \\
\end{tabular}
\end{ruledtabular}
\end{table}
